\documentstyle[12pt,rei2]{article}
\begin{document}


\title{Reionization of the Universe and the Early Production of Metals}
\author{Nickolay Y.\ Gnedin\altaffilmark{1} and
Jeremiah P.\ Ostriker\altaffilmark{2}}
\altaffiltext{1}{University of California, Berkeley Astronomy Department,
Berkeley, CA 94720; e-mail: \it gnedin@astron.berkeley.edu}
\altaffiltext{2}{Princeton University Observatory, Peyton Hall, 
Princeton, NJ 08544; e-mail: \it jpo@astro.princeton.edu} 


\load{\scriptsize}{\sc}

\def\A{{\cal A}}
\def\B{{\cal B}}
\def\ion#1#2{\rm #1\,\sc #2}
\def\HI{{\ion{H}{i}}}
\def\HII{{\ion{H}{ii}}}
\def\GI{{\ion{He}{i}}}
\def\GII{{\ion{He}{ii}}}
\def\GIII{{\ion{He}{iii}}}
\def\MH{{{\rm H}_2}}
\def\Hp{{{\rm H}_2^+}}
\def\Hm{{{\rm H}^-}}

\def\dim#1{\mbox{\,#1}}

\def\figdir{.}
\def\placefig#1{#1}

\def\capTF{
Transfer functions for the dark matter ({\it solid line\/}) and the
baryons ({\it dotted line\/}) at $z=100$ in ratio to the BBKS transfer function
for the CDM+$\Lambda$ cosmological model with $\Omega_0=0.35$, $h=0.7$, and
$\Omega_b=0.03$.
}

\def\capSF{
{\it Upper panel\/}: the fraction $f_*$ os the total baryonic mass 
locked into stars as a function of redshift for run B. 
{\it Middle panel\/}: the
comoving star formation rate in units of $\dim{M}_\odot/\dim{yr}/(h^{-1}
\dim{Mpc})^3$ ({\it solid line\/})
and the comoving stellar feedback rate, normalized to the same units
({\it dotted line\/}),
as a function of redshift for the same run.
{\it Lower panel\/}: the gas clumping $C_{bb}\equiv\langle\rho^2\rangle/
\langle\rho\rangle^2$ ({\it solid line\/}) and the ionized hydrogen clumping
$C_\HII$ (discused in \S 3.1; {\it dotted line\/})
as a function for redshift for the same run.
}

\def\capTI{
Evolution of the radiation intensity $J_{21}$ ({\it upper panel\/}),
average neutral hydrogen fraction $f_{\HI}$ ({\it middle panel\/}), and the
average temperature ({\it lower panel\/}) as a function of redshift
for run B. Solid lines show the volume average and dotted lines show the
mass average for the temperature and neutral hydrogen fraction.
}

\def\capDA{
{\it Upper panel\/}:
continuum suppression factors $1-D_A$ as a function of redshift for Run B
({\it solid line\/}) together with the data from Jenkins \& Ostriker (1991)
({\it filled circles\/}).
{\it Lower panel\/}: 
the Gunn-Peterson optical depth as a function of redshift for Run B 
({\it solid line\/}). Shown as a solid triangle is the observational upper
limit from Jenkins \& Ostriker (1991) and as an open circle
is the upper limit from Giallongo et al.\ (1994).
}

\def\capFZ{
The mass weighted ({\it solid line\/}) and the volume weighted ({\it dotted 
line\/}) fraction of the intergalactic gas that was processed
in stars as a function of redshift. The dashed line shows 
$\rho^{5/3}$ weighting which roughly corresponds to the weighting by the
column density of a Lyman-alpha absorber.
The right hand scale corresponds to the normalization
with the adopted value for the yield $y=0.020$.
}

\def\capMF{
The mass- ({\it upper panel\/}) and volume- ({\it lower panel\/}) weighted
probability distribution to find a fluid element with given values of the
fraction of metal enriched gas and the total gas density at $z=4$. 
The right hand scale corresponds to the normalization
with the adopted value for the yield $y=0.020$. The bold line shows the average
probability, solid contours mark the probability above average, and the dashed
contours show the probability below average; contour spacing is logarithmic
with the increment of $1/3$.
}

\def\capZD{
The average fraction of the metal enriched gas as a function of total gas 
density at four different redshifts $z=4$, $5$, $6$, and $8$. 
The right hand scale corresponds to the normalization
with the adopted value for the yield $y=0.020$.
}
 
\def\capZE{
Distributions of the dark matter ({\it left column\/}), 
the cosmic gas ({\it middle column\/}), and the stars ({\it right column\/}) 
at three different redshifts: $z=5.4$ ({\it upper row\/}), $z=4.9$
({\it middle row\/}), and $z=4.3$ ({\it lower row\/}) represented by particles
in a thin slice with the width of $12h^{-1}$ comoving kpc. The square and 
the circle mark two merging objects, and the oval pointed to by an arrow
marks ejection of the metal enriched gas from the merger product.}

\def\capEM{
The spectrum of the $21\dim{cm}$ emission from the intergalactic gas
at high redshift ({\it solid line\/}). 
The sharp break in the spectrum corresponds to the redshift
of reionization. The two dashed lines show the expected rms fluctuations
in this radiation on the 1 ({\it upper curve\/}) and 10 ({\it lower curve\/})
arcmin angular scale.
Shown with the dotted line is the Cosmic Microwave
Background radiation intensity reduced by a factor of a hundred to fit
on the same scale.
}

\def\capEF{
The fine structure of the $21\dim{cm}$ emission at three
different redshifts. Each spectrum is centered at the reshifted frequency
of the $21\dim{cm}$. The emission terminates at high velocity values due to 
the final size of the simulation box.}

\def\capSS{
The spectra of the background radiation at three different 
redshifts. Ionization edges corresponding to the hydrogen and helium
ionization are well observed at all times.}

\def\capZM{
The fraction of metal enriched gas as a function of the total mass for
all bound objects with $M_{\rm tot}>10^8h^{-1}M_{\sun}$ and 
$M_*>10^5h^{-1}M_{\sun}$ at $z=9.3$ ({\it solid squares\/}),
$z=5.9$ ({\it stars\/}), and $z=4.3$ ({\it open circles\/}).
The right hand scale corresponds to the normalization
with the adopted value for the yield $y=0.020$.
}

\def\capFM{
The fraction of stars in the total mass ({\it a\/}) and in the total baryonic
mass ({\it b\/}) as a function of the total mass of an object for all massive
objects at three different redshifts as explained before.}

\def\capMV{
The total mass as a function of the dark matter velocity dispersion 
({\it a\/}) and the stellar mass as a function of stellar velocity
dispersion ({\it b\/}) for all massive
objects at three different redshifts as explained before. The solid lines
have a slope of 2 in ({\it a\/}) and 4 in ({\it b\/}).}

\def\tableone{
\begin{table}
\caption{Numerical Parameters}
\medskip
$$
\begin{tabular}{cccccc}
Run & $N$ & Box size & Total mass res.\
 & Spatial res.\ & Dyn.\ range \\ \tableline
A & $128^3$ & $2h^{-1}{\rm\,Mpc}$ & $3.7\times10^5h^{-1}\dim{M}_{\sun}$ & $1.0h^{-1}{\rm\,kpc}$ & $2000$ \\
B & $ 64^3$ & $2h^{-1}{\rm\,Mpc}$ & $2.9\times10^6h^{-1}\dim{M}_{\sun}$ & $3.0h^{-1}{\rm\,kpc}$ & $640$ \\
C & $ 64^3$ & $1h^{-1}{\rm\,Mpc}$ & $3.7\times10^5h^{-1}\dim{M}_{\sun}$ & $1.5h^{-1}{\rm\,kpc}$ & $640$ \\
\end{tabular}
$$
\end{table}
}

\def\tabletwo{
\begin{table}
\caption{Molecular Hydrogen Chemistry Rates}\label{tabmolhyd}
\medskip
$$
\begin{tabular}{rcl}
No.\tablenotemark{a} & Reaction & Rate notation \\ \tableline
 7 & $\HI + e    \rightarrow \Hm        $ & $ k_7 $ \\
 8 & $\HI + \Hm  \rightarrow \MH + e    $ & $ k_8 $ \\
 9 & $\HI + \HII \rightarrow \Hp        $ & $ k_9 $ \\
10 & $\HI + \Hp  \rightarrow \MH + \HII $ & $ k_{10} $ \\
11 & $\HI + \MH  \rightarrow 3\HI       $ & $ k_{11} $ \\
12 & $\MH + \HII \rightarrow \HI + \Hp  $ & $ k_{12} $ \\
13 & $\MH + e    \rightarrow \HI + \Hm  $ & $ k_{13} $ \\
14 & $\MH + e    \rightarrow 2\HI + e   $ & $ k_{14} $ \\
15 & $\MH + \MH  \rightarrow \MH + 2\HI $ & $ k_{15} $ \\
16 & $\Hm + e    \rightarrow \HI + 2e   $ & $ k_{16} $ \\
17 & $\Hm + \HI  \rightarrow 2\HI + e   $ & $ k_{17} $ \\
18 & $\Hm + \HII \rightarrow 2\HI       $ & $ k_{18} $ \\
19 & $\Hm + \HII \rightarrow \Hp + e    $ & $ k_{19} $ \\
20 & $\Hp + e    \rightarrow 2\HI       $ & $ k_{20} $ \\
21 & $\Hp + \Hm  \rightarrow \MH + \HI  $ & $ k_{21} $ \\
27 & $\Hm + \gamma \rightarrow \HI + e    $ & $ k_{27} $ \\
28 & $\Hp + \gamma \rightarrow \HI + \HII $ & $ k_{28} $ \\
29 & $\MH + \gamma \rightarrow \Hp + e    $ & $ k_{29} $ \\
30 & $\Hp + \gamma \rightarrow 2\HII + e  $ & $ k_{30} $ \\
31 & $\MH + \gamma \rightarrow 2\HI       $ & $ k_{31} $ \\
\tablenotetext{a}{as in Shapiro \& Kang\markcite{SK87} 1987}
\end{tabular}
$$
\end{table}
}

\begin{abstract}

We simulate a plausible cosmological model in considerable physical and
numerical detail through the successive phases of reheating (at $10\la z
\la20$), formation of Pop III stars at $z\sim15$ (due to $\MH$ cooling),
with subsequent reionization at $z\approx7$. We assume an efficiency of
high mass star formation appropriate to leave the universe, after it becomes
transparent, with an ionizing background $J_{21}\approx0.4$ (at $z=4$), 
near (and
perhaps slightly below) the observed value. Since the same stars produce the
ionizing radiation and the first generation of heavy elements, a mean 
metallicity of $\langle Z/Z_{\sun}\rangle\sim1/200$ is produced in this
early phase, but there is a large variation about this mean, with the high 
density regions having $Z/Z_{\sun}\approx1/30$ and low density regions
(or the Lyman-alpha forest with $N_{\HI}\la10^{13.5}\dim{cm}^2$) having
essentially no metals. 

Reionization, when it occurs, is very rapid
(phase change-like), which will leave a signature which may be
detectable by very
large area meter-wavelength radio instruments. Also, the background 
UV radiation
field will show a sharp drop of $\sim10^{-3}$ from $1\dim{Ryd}$ to
$4\dim{Ryd}$ due to absorption edges. 

The simulated volume is too small to form $L_*$ galaxies, but the smaller
objects which are found in the simulation obey the Faber-Jackson relation.

In order to explore theoretically this domain of ``the end of the
dark ages'' quantitatively, numerical simulations must have a mass
resolution of the order of $10^{4.5}\dim{M}_{\sun}$ in baryons, high spatial
resolution ($\la1\dim{kpc}$ to resolve strong clumping, and allow for
detailed and accurate treatment of both the radiation field and 
atomic/molecular physics.

\end{abstract}


\section{Introduction}

In a previous paper (Ostriker \& Gnedin 1996; Paper I) we showed, for a
specific cosmological model, how an early generation of stars would be
expected to form from the first nonlinear self-gravitating clumps, which
developed while the universe was still relatively cold, and was
primarily filled
with neutral gas. This first generation, ``Population III'', which would 
reside in widely and fairly smoothly distributed clumps of stars somewhat more
massive than globular clusters, will reheat and reionize the universe, leading 
to an end of what Martin Rees has called the ``Dark Ages'', during the
interval $15>z>7$. Since the same stars produce both ionizing photons during
their main sequence phase and metals when they end their lives as type II 
supernovae, this early phase of the universe, ``the clearing of the fog'', 
also and inevitably leaves an irregular contamination with metals at the
low (average) level of $\langle Z\rangle/Z_\odot\sim10^{-2.5}$.

In the current paper we provide the physical basis for the conclusions reached
in Paper I, and also address a number of related questions, among them being
the following:
\begin{enumerate}
\item When, in terms of the Gunn-Peterson effect (Gunn \& 
Peterson\markcite{GP65} 1965), should the universe have become transparent?
And how closely are the processes labeled ``reheating'' and
``reionization'' related?
\item How would one characterize the spatial variations in the metal abundance
produced by the early generation of stars?
\item What was the degree of clumpiness in the gas during these early phases?
(This is a question, typically neglected in most of previous treatments.)
\item Will specially designed large area telescopes looking at the (rest 
frame) $21\dim{cm}$ radio sky be able to detect this early development of
structure?
\item How important were the various cooling processes (atomic, molecular, 
and dust), the various heating processes (gravitational collapse, UV photons,
supernovae), and radiative shielding of high density regions (by gas or dust)?
\item What were the properties of the stellar groups formed at high
redshifts? Are they observable?
\end{enumerate}

These and other questions are addressed in the context of a specific 
cosmological model. In a final section we speculate on how robust our
conclusions are to variations of the assumed model.

In discussing various physical effects that play role at the end of
the ``Dark Ages'', we must note the many authors who
have studied these processes before, from the pioneering work of
Couchman \& Rees\markcite{CR86} (1986) and Shapiro \& Giroux\markcite{SG87}
(1987) to the recent careful investigations by
Shapiro, Giroux, \& Babul\markcite{SGB94} (1994),
Tegmark and Silk\markcite{TS94,95} (1994, 1995), Kawasaki \&
Fukugita\markcite{KF94} (1994), Shapiro\markcite{S95} (1995),
Giroux \& Shapiro\markcite{GS} (1996),
Tegmark et al.\markcite{Tea96} (1996) to the one-dimensional simulations
by Haiman, Thoul, \& Loeb\markcite{HTL95} (1995) and Haiman \& 
Loeb\markcite{HL96} (1996).
While varying amounts of physically detailed modelling were included
in all of those papers, there was a critical factor that could not be
followed in any of the semi-analytic treatments: the clumping of the gas.
However, {\it all\/} the relevant processes are dependent on the 
clumping factor
$C_{bb}\equiv\langle\rho^2\rangle/\langle\rho\rangle^2$. Included among these
processes are {\it recombination, cooling, gravitational collapse, molecular 
hydrogen formation\/} and numerous others. 
Whereas one-dimensional simulations
are able to account for the clumping around a single spherically symmetric
object, they must neglect the clumping within the flow and therefore they
cannot predict the average clumping of the gas in the universe,
and are not able to follow the evolution of the radiation field
in any reasonable detail.
Thus, even the qualitative conclusions of the previous papers may be
in doubt until they are confirmed by a fully nonlinear three dimensional 
threatment. A legitimate question to ask at this point is the following:
Does the present simulation, reported on in this paper have sufficient
spatial resolution to correctly compute the clumping? The answer is
either ``yes'' or ``barely enough'', depending on ones degree of
optimism. On scales below the Jeans mass clumping should be relatively
unimportant as it cannot (by definition of the Jeans mass) be assisted
by gravity. At a characteristic temperature of $10,000\dim{K}$, the
Jeans mass is around $10^{10}\dim{M}_{\sun}$ at $z=4$ for the average
density of the universe, and is around $10^8\dim{M}_{\sun}$ for the
overdensity of $10^4$ (which is reliably resolved by our simulations),
whereas our nominal total mass resolution is $10^{5.5}\dim{M}_{\sun}$
(cf Table 1). Numerical experiments that we have performed indicate that
the resolution needed to permit the existence of a first (Pop III) generation
of stars is of the order of $10^{5.5}\dim{M}_{\sun}$, which we just achieve
in the computations reported on here. Work currently underway should allow
us to push the limit to a comoving spatial scale of $\approx0.5h^{-1}
\dim{kpc}$ and a mass scale of $1.0\times10^5\dim{M}_{\sun}$, while 
simultaneously allowing better for the missing large-scale
power.

\section{Method}

\subsection{Physical Ingredients of a Numerical Simulation}

In order to adequately simulate to at least moderate accuracy
the evolution of the
universe at high redshift, we must allow for all of the most important
physical 
processes. Fortunately, by restricting ourselves to early stages of formation
of structure, we can incorporate essentially all physics which affects
formation of structure at moderate densities ($n\lesssim1\dim{cm}^{-3}$), 
contrary to simulations of
galaxy and large-scale structure formation at low redshifts, which are
generically missing important physical ingredients. The main reason behind
this difference between simulating high and low redshift universe is that
most of poorly understood complicated physical processes, i.e.\ effects
of heavy elements, dust, magnetic fields, are thought to be less important
at high redshift, where most of the intergalactic gas is of 
essentially primeval composition and magnetic fields have not yet had
time to build up to a significant strength (cf Kulsrud et 
al.\markcite{KCOR96} 1996).

We have therefore undertaken to simulate numerically the evolution
of the intergalactic medium at high redshift, including essentially all
physical processes that we identified as important for the high redshift
evolution of the universe. We have used the SLH-P$^3$M cosmological 
hydrodynamic
code as described by Gnedin\markcite{G95} (1995) and 
Gnedin \& Bertschinger\markcite{GB96} (1996) to 
follow the evolution
of the dark matter and the cosmological gas with high Lagrangian resolution.
We have also included detailed atomic and molecular physics of a gas of
primeval composition, following in non-equilibrium time-dependent fashion
the ionization and recombination of hydrogen and helium as well as
formation and destruction of hydrogen molecules in the ambient 
radiation field. The radiation field, in turn, allows for sources
of radiation (quasars and massive stars), sinks (due to 
continuum opacities) and cosmological effects. 
In particular, we have taken into account the fact that dense lumps will be
shielded from the background radiation field. This reduces the heating rates
for dense clumps and makes it nearly certain that once they have formed
and started to collapse, the process will be irreversible. They will continue
to collapse and fragment even though outer, lower density regions may
reheat and be stabilized against further collapse. However, since the precise
calculation of the shielding effect requires carrying on the whole radiative
transfer in the highly inhomogeneous medium, which is well beyond the
capabilities of existing computers, we adopted a simple approximation which
we call the Local Optical Depth approximation. In its essentials this 
approximation treats absorption as localized but the sources of radiation as
smoothly distributed. We derive radiative transfer 
equations in the expanding universe in the Appendix A, and then introduce
the Local Optical Depth approximation in the Appendix B. In Appendix C
we describe in detail our method of following formation and destruction of
molecular hydrogen.

The ionizing radiation is emitted by stars and quasars. However, it is
impossible to simulate star formation within the cosmological framework 
directly, and we therefore are forced to use a phenomenological approach
to account for star formation and feedback processes. 
In regions which are cooling
and collapsing we have allowed the formation of point-like ``stellar''
subunits according to the Cen \& Ostriker\markcite{CO92} (1992) 
algorithm as adapted to the
SLH code by Gnedin\markcite{G96} (1996). 
However, we introduce the following change to the
Cen \& Ostriker algorithm. In their original paper Cen \& Ostriker introduced
as a numerical convenience
the density cut-off as a value below which no formation of stellar subunits
was permitted. The value of the cutoff is arbitrary, and final results
are somewhat dependent on its value, as has been shown by 
Gnedin\markcite{G96} (1996) and
Katz, Weinberg, \& Hernquist\markcite{KWH} (1996). In our improved version 
of the star formation algorithm, we abandon the density cut-off value in 
the belief that as long as the code has not reached its resolution limit,
the evolution of the gas is followed accurately, and there is no need to
introduce this phenomenology altogether. 
Therefore, we allow for star 
formation only in the regions which have reached the resolution limit, since,
in those regions, the further collapse of the gas cannot be followed accurately
and stellar particles {\it must} be created in order to allow the gas
to sink beyond the resolution limit of a simulation. We
identify this limit as the limit when the Lagrangian behavior of 
the SLH code switches
to the Eulerian one. Numerically, this corresponds to places where the
maximum eigenvalue of the softening tensor $\sigma_{ij}$ reaches $1/2$ 
(see Gnedin \& Bertschinger\markcite{GB96} [1996] for the definition of the
softening tensor $\sigma_{ij}$), and physically corresponds to the spatial
resolution shown in Table 1 for the various runs we have performed.
These numbers are essentially our only free parameters. The sum of them is
fixed by matching the background radiation field at $z=4$ and the ratio
determines the softness of the ionizing background spectrum 
(cf Fig.\ \ref{figSS}), also an observable.

As soon as a stellar particle forms, it releases radiation and (in proportion)
metal rich gas, which we have considered in the treatment of cooling. Since
we cannot distinguish between stars and quasars in our treatment of star
formation, we allow each stellar particle to emit a composite spectrum of 
radiation, consisting of mixture of both the quasar and young star spectra.
The shape of the spectrum is shown in Fig.\ 1 of Gnedin\markcite{G96} (1996).
The amount of radiation emitted is again a free parameter of the
phenomenological star formation algorithm, and we choose it so that we
produce ionizing radiation at low redshift in the amount which is close to the
observed value.\footnote{However, since we cannot determine the final
value for the ionizing intensity until after we have run the simulation,
we can only approximately reproduce the observational value for the
ionizing intensity.} We assume an efficiency $\varepsilon_{\rm UV,S}=
6\times10^{-5}$ for the stellar component of the source spectrum and the
equal value of $\varepsilon_{\rm UV,Q}=6\times10^{-5}$ for the QSO-like
component of the source spectrum. These numbers are essentially our only free
parameters. The sum of them is fixed by matching the background radiation 
field at $z=4$, and the ratio determined the softness of the ionizing
bacground (cf Fig.\ \ref{figEM}), also an observable.

We have also allowed for the injection of the thermal energy from supernova
explosions into the intergalactic gas according to Cen \& Ostriker (1992)
methodology. However, at the resolution we achieve
in the simulations reported here the gas density is so high that all this
thermal energy is immediately radiated away producing negligible net effect
on the evolution of the universe. This results stems purely from the
inadequate treatment of the supernova explosions since gas inside the
supernova bubbles is too hot to cool. Therefore, the appropriate treatment
of the supernova explosions would require following hydrodynamics
of the multi-phase medium (Cowie, McKee, \& Ostriker 1981)
and modelling interactions between the phases,
which is beyond the frame of this work. We, therefore, emphasize that while
supernova explosions are formally incorporated in our simulations, they are
not followed adequately and, therefore, their effects may be significantly
underestimated in our treatment.

The only important piece of physics, that we are totally
missing in our simulations,
is the nonuniformity of the sources of radiation (as explained in Appendix B).
One, therefore, should bear this in mind when interpreting the results
of our simulations. However, since we compute all average quantities
as appropriate, we expect that the evolution of the universe as whole,
and, in particular, the evolution of average quantities, is reproduced
adequately by our simulations.

\subsection{Cosmological Models}

\placefig{
\begin{figure}
\insertfigure{\figdir/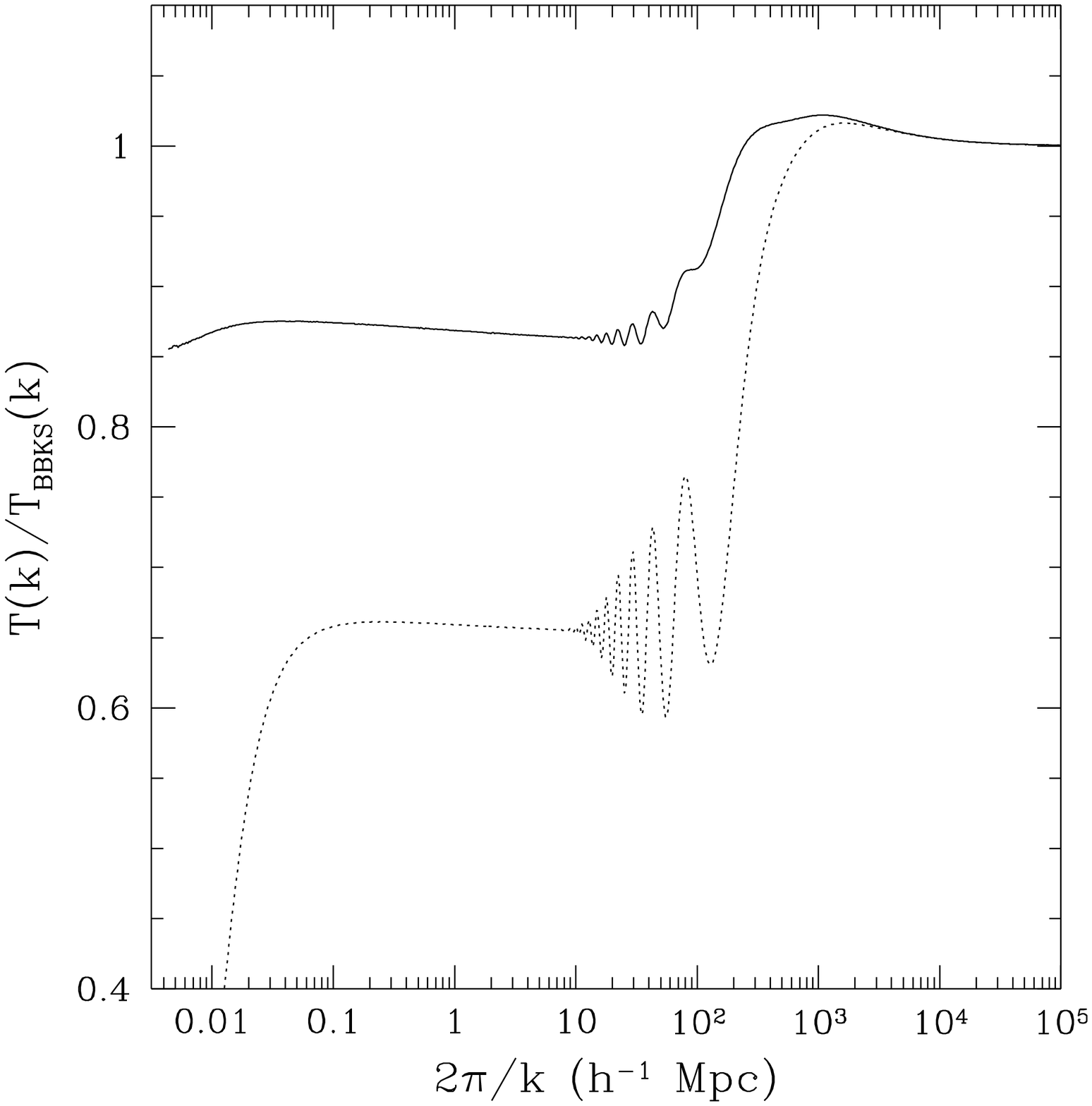}
\caption{\label{figTF}\capTF}
\end{figure}
}
We have adopted a CDM+$\Lambda$ cosmological model as a framework for
our investigations. However, we believe that qualitative results of our
simulation are applicable to the whole family of the CDM-type models,
and we emphasize those qualitative conclusions together with specific
quantitative results for the model we use. We fix the cosmological parameters
as follows:
$$
\begin{array}{rcl}
	\Omega_0 & = & 0.35, \\
	       h & = & 0.70, \\
	\Omega_b & = & 0.03, \\
\end{array}
$$
which is close to the ``concordance'' model of Ostriker \& 
Steinhardt\markcite{OS95} (1995).
We normalize it to {\it COBE} $10^o$ measurement and allow for a small
tilt with $n=0.96$. This gives the value of $\sigma_8=0.67$ for the power
spectrum normalization at $8h^{-1}\dim{Mpc}$ at $z=0$. In order to compute
the initial conditions accurately, we computed the linear transfer functions
for the model using the linear Boltzmann code similar to (but different from) 
the COSMICS package (Bertschinger\markcite{B95} 1995). We show in Fig.\ \ref{figTF}
transfer functions for both the dark matter (solid line) and the baryons
(the dotted line) in ratio to the BBKS transfer functions
(Bardeen et al.\markcite{BBKS86} 1986) at $z=100$, where we start our
simulations.

We have performed three runs with different box sizes and numerical resolution
to assess the importance of different scales and estimate the uncertainty due
to the finite resolution of our simulations. The adopted parameters of those
runs are compiled in Table 1. All runs were stopped at $z=4$ since at that
time the rms density fluctuation at $2h^{-1}\dim{Mpc}$ scales is 0.4 and
the absence of nonlinear wavelengths larger than the box size 
renders the simulations senseless.
\placefig{\tableone}
The largest of our simulations,
run A, is our fiducial run against which to compare results of 
the two smaller runs.
However, it happened by accident that the particular realization of initial
conditions we used for run A had substantially less power on large scales
($1-2h^{-1}{\rm\,Mpc}$) than the true power spectrum, and we have to take 
this fact into account when interpreting our results. In particular,
this affects the precise value for the redshift
of reionization and the evolution at late times, but at earlier times,
$z\ga10$, when the density correlation length is much smaller than the
box size, the absence of large waves has little effect.

\section{Results}

\subsection{Reheating and Reionization}

It is often assumed in the current literature that reheating and reionization
of the universe at high redshift are two sides of the same physical process, 
and even the terms ``reheating'' and ``reionization'' are sometimes
used interchangeably. In this section we demonstrate that while the physical
cause of both of those processes is the same: the high energy radiation field,
they are nevertheless two different physical processes, separated in time
as well as proceeding at different speeds. 

\placefig{
\epsscale{0.65}
\begin{figure}
\insertfigure{\figdir/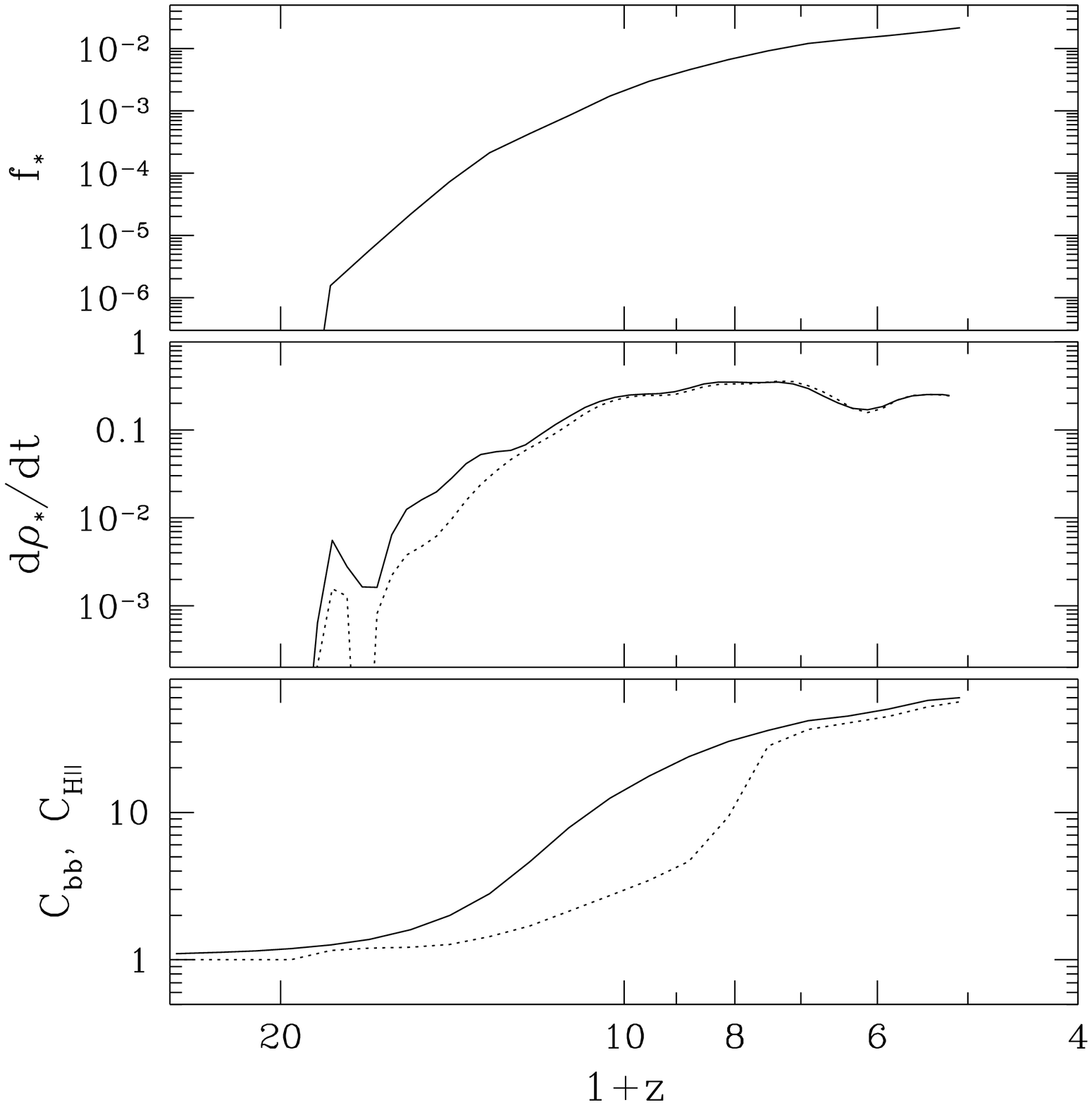}
\caption{\label{figSF}\capSF}
\end{figure}
}
Since it is the stars that produce the ionizing radiation, we first show in
Figure \ref{figSF} the fraction $f_*$ of the total baryonic mass which 
is locked in stars (the upper panel) and the comoving star formation rate 
$d\rho_*/dt$ in units of $h^3\dim{M}_{\sun}/\dim{yr}/\dim{Mpc}^3$,
as
a function of redshift (the middle panel) from out run B. 
Also in the middle panel we show with the
dotted line the comoving feedback rate, i.e.\ the rate of production of
high energy photons and metals (for convenience, we convert it to the same 
units). The feedback effects are delayed after the
formation of a stellar particle according to
the methodology of Cen \& Ostriker\markcite{CO92} (1992; see 
Gnedin\markcite{G96} 1996 for details).
Note the existence of the first peak of Pop III
star formation at $z\approx16$ in otherwise smooth star formation rate.
The slow down of the star formation at $z<6$ is due to the lack of large-scale
waves in our simulation box.

\placefig{
\begin{figure}
\insertfigure{\figdir/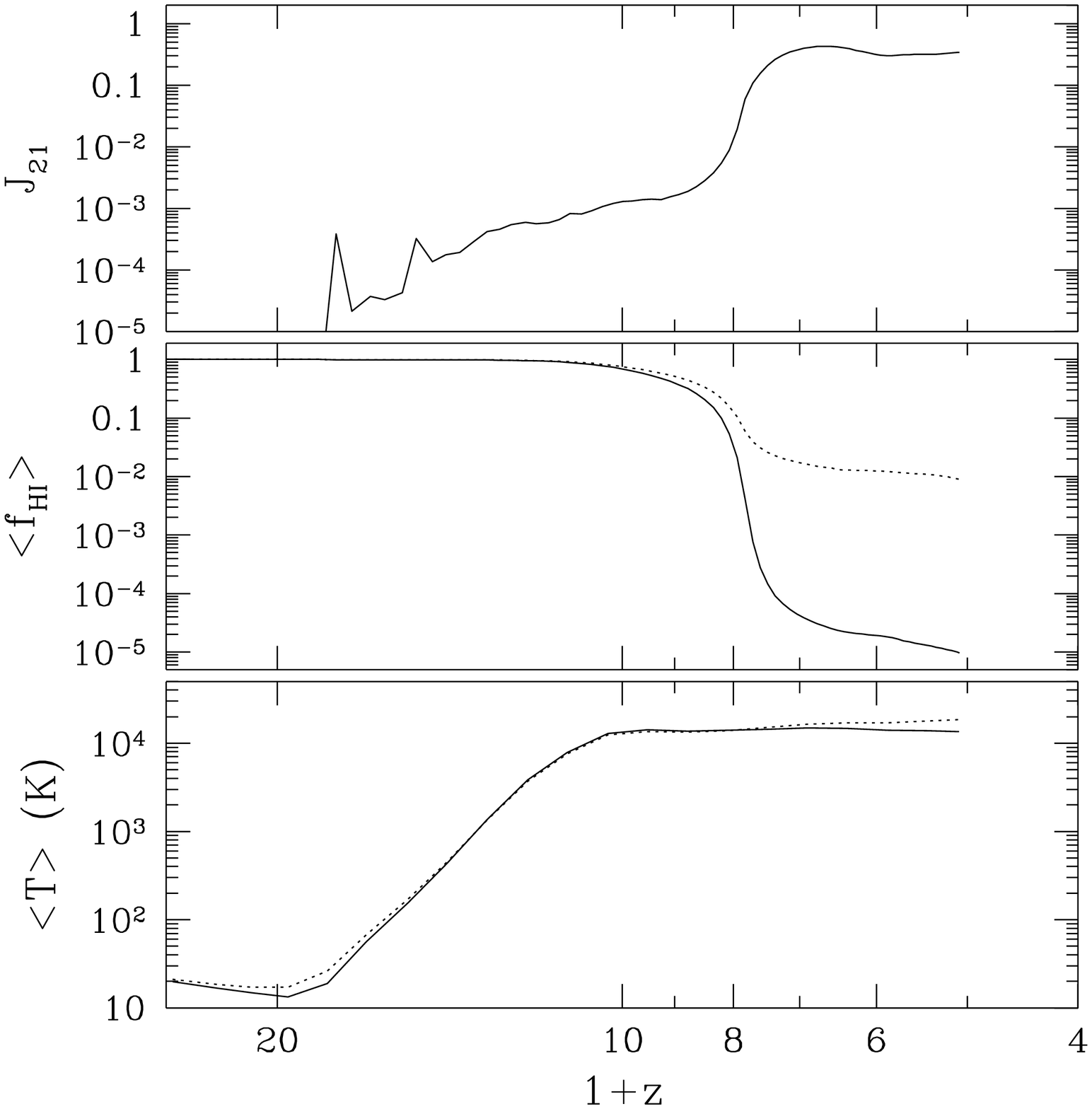}
\caption{\label{figTI}\capTI}
\end{figure}
}
Fig.\ \ref{figTI} shows the temporal evolution for the gas temperature
(lower panel) and the neutral hydrogen fraction (middle panel). We also
show the evolution of the ionizing radiation intensity as measured by the
quantity $J_{21}$, called the ionizing intensity, defined as follows:
\begin{equation}
	J_{21} \equiv 
	{\int J_\nu \sigma_\nu d\nu/\nu\over \int \sigma_\nu d\nu/\nu}
	\times
	{1\over10^{-21}\dim{erg}\dim{cm}^{-2}\dim{s}^{-1}\dim{Hz}^{-1}
	\dim{sr}^{-1}},
	\label{J21def}
\end{equation}
where $J_\nu$ is the radiation intensity and $\sigma_\nu$ is the hydrogen
photoionization cross-section. The observed values for $J_{21}$ lie between
$0.3$ and $1.5$ (Savaglio et al.\markcite{SCDFGM96} 1996; Cooke, Espey, \& 
Carswell\markcite{CEC96} 1996) at $z\sim3-4$.

We immediately note that there is a relatively long epoch from $z=20$ to
$z=10$, encompassing many Hubble times at that redshift,
when the volume and mass weighted average temperature of 
the universe increases from the value
$T=11\dim{K}$ at $z=20$, dictated by the pure adiabatic expansion of the
universe, to the value $T=14,000\dim{K}$, fixed by the atomic physics.
This increase in temperature, or {\it reheating\/} of the universe,
is accompanied by a slow increase in the value of ionizing intensity
up to $J_{21}\sim10^{-3}$, but the universe stays essentially neutral, with
the neutral hydrogen fraction decreasing slowly until $z=7$, when suddenly
the neutral fraction drops by several orders of magnitude at essentially
constant temperature and the ionizing
intensity $J_{21}$ increases to a value of around $0.5-1.0$. This moment,
which happens in a small fraction of a unit redshift (i.e.\ in a small
fraction of the then Hubble time), is called {\it reionization\/} of the 
universe and occurs almost as a phase change.

We immediately conclude that reheating and reionization are two different
physical processes; reheating is a slow process happening at high redshift,
$z>10$, followed by a sudden reionization at a lower redshift. During
reheating the value of the ionizing intensity stays below $J_{21}<10^{-3}$,
because the universe remains mainly neutral at that time and the absorption
is large. In order to understand why the absorption of ionizing photons is
significant, let us estimate the recombination time, i.e.\ the time a
hydrogen ion needs to recombine after being ionized by a high-energy
photon. For the temperature $T=10^4\dim{K}$ the value of the recombination
coefficient $R$ is $4.2\times10^{-13}\dim{cm}^3\dim{s}^{-1}$ assuming case A
recombination. The equation for the ionized hydrogen number density can be
written as follows:
\begin{equation}
	\dot{n}_\HII = -3Hn_\HII + n_\HI\Gamma - R n_e n_\HII,
	\label{hiieq}
\end{equation}
where $H$ is the Hubble constant, $\Gamma\equiv4.3\times10^{-12}J_{21}
\dim{s}^{-1}$ is the photoionization rate, and $n_e$ is the free electron 
number
density. Equation (\ref{hiieq}) holds at every fluid element in the universe. 
In order to compute the average number density of the ionized hydrogen, we 
have to average equation (\ref{hiieq}) over the volume of the universe,
obtaining:
\begin{equation}
	\dot{\bar{n}}_\HII = -3H\bar{n}_\HII + \langle n_\HI\Gamma\rangle -
	\langle R n_e n_\HII\rangle,
\end{equation}
where the bar symbol means the volume average.
Assuming, that $R$ is constant in space and that helium is not
ionized (i.e.\ $n_e=n_\HII$), we can obtain the following expression
for the recombination time:
\begin{equation}
	t_R \equiv {\bar{n}_\HII\over R \langle n_\HII^2 \rangle},
	\label{trone}
\end{equation}
or for the ratio of the recombination time to the Hubble time, $t_H=1/H$:
\begin{equation}
	{t_R\over t_H} = {H\over \bar{n}_\HII R C_\HII} =
	67{a^{3/2} h\sqrt{\Omega_0}\over x C_\HII},
	\label{trtoth}
\end{equation}
where $a$ is the scale factor, $x$ is the volume average ionization fraction,
$x\equiv \bar{n}_\HII/\bar{n}_{\rm H}$, 
and $C_\HII$ is the ionized hydrogen clumping
factor, $C_\HII\equiv\langle n_\HII^2\rangle/\langle n_\HII\rangle^2$,
shown in the lower panel of Fig.\ \ref{figSF} with the dotted line, 
which is somewhat lower than the total gas clumping factor
$C_{bb}\equiv\langle\rho^2\rangle/\langle\rho\rangle^2$ because 
higher density regions are typically less ionized.
For the model under consideration, $h\sqrt{\Omega_0}=0.7\sqrt{0.35}=0.41$,
and for the fully ionized gas, $x=1$, with no clumping, the recombination time
is equal the Hubble time at the redshift $z_{eq}=8$, which implies that
if the universe stayed uniform, once it was ionized after $z=8$, it will
not be able to recombine ($t_R/t_H>1$). This is a well known result.
However, since the universe is in fact strongly clumped, and
the clumping factor $C_\HII\gg1$, the absorption is much higher than
that for the uniform universe. In particular, the clumping factor $C_\HII$
is equal 12 at $z=8$ (and it is even larger in the real
universe since we underestimate clumping due to the finite resolution of
our simulations), and the recombination time is equal to the Hubble time at
$z=8$ for the ionization fraction as low as $x=1/C_\HII=0.08$. This means that
it is sufficient to have only 8\% of all hydrogen ionized in order to be
able to absorb at least one ionizing photon per Hubble time per hydrogen
atom (the corresponding number is 10\% for the recombination case B).

It is also easy to understand why the reheating phase should precede 
the reionization phase. Let us consider the ratio of the photoheating time,
$$
	t_{\rm heat}\equiv{3/2k_BT\over E_J\Gamma},
$$
to the photoionization time,
$$
	t_{\rm ion}\equiv{1\over\Gamma},
$$
where $k_B$ is the Boltzmann constant, and
$$
	E_J \equiv
	{\int_{h\nu_0}^\infty (h\nu-h\nu_0)\sigma_\nu J_\nu d\nu/\nu \over
	\int_{h\nu_0}^\infty \sigma_\nu J_\nu d\nu/\nu},
$$
is the average energy a hydrogen atom receives in act of ionization.
For our case, $E_J=k_B9.3\times10^4\dim{K}$ and $T=10^4\dim{K}$
at $z=10$, and
\begin{equation}
	{t_{\rm heat}\over t_{\rm ion}}={3\over2}{k_BT\over E_J}=0.16
	\label{rrnum}
\end{equation}
which implies that the radiation background is much more (by a factor of 6)
efficient in
heating the gas at high redshift than in ionizing it. The reason is simply 
that, unless the ionizing radiation field is very soft, excess heat is
liberated for each act of photoionization.

Finally, we can understand why the reionization is so rapid. Since the
recombination time is short compared to the Hubble time, 
there is an approximate balance between the
emission of ionizing photons from the sources and the absorption 
(by ionizing a hydrogen atom which subsequently recombines). If the
emission increases, so does the ionization fraction $x$, the recombination
time decreases, and the absorption increases; inversely, the decrease in the
emission leads to the subsequent decrease in the ionization fraction and
an increase in the recombination time. Since the star formation rate is 
increasing with time, as shown in the middle panel of Fig.\ \ref{figSF},
driven by the further collapse of hierarchical structure, the ionization
fraction slowly increasing in response to the increase in the emission
of ionizing photons, until it reaches values close to unity. At this
point a further increase in the recombination rate cannot be achieved,
emission and absorption get out of balance, and the universe ionizes
at the time scale of the emission, which is at the point of loss of balance
is approximately equal to the recombination time. We, therefore, conclude
that the sharp decrease in the neutral hydrogen fraction from the values
about $0.1$ at the beginning of (complete) reionization to final values of the
order of $10^{-5}$ is achieved on a time-scale of
\[
	t_{\rm rei} \sim t_R={t_H\over C_\HII}\left(1+z_{\rm rei}
	\over9\right)^{-3/2},
\]
which for our case of $z_{\rm rei}=7$ has a value of
\[
	t_{\rm rei} \sim {t_H\over 10},
\]
in excellent agreement with the actual reionization time measured from 
Fig.\ \ref{figTI}. This effect of fast reionization is entirely due to
the high value of the clumping factor. Were the clumping ignored in our
consideration, we would conclude that the reionization time is more than
10 times larger than the actual value is, and that reionization is a
slow process continuing for more than one Hubble time
(as long as the source evolution is also slow as in Fig.\ \ref{figSF}).
This is precisely the conclusion achieved in some of the previous 
work on the reionization
where the clumping was ignored 
(see, for example, Haiman \& Loeb\markcite{HL96} 1996).

We point out here
that the specifics of the cosmological model does not enter this
conclusion and the prediction that slow reheating happens first, followed
later by sudden reionization, is a generic prediction for any cosmological
model (with UV photons providing both heat and ionization), 
not necessarily a CDM-type one.

\subsection{The Gunn-Peterson Effect in the Intergalactic Medium}

The complete reionization of the universe manifests itself in the 
absence of the
Gunn-Peterson absorption trough (Gunn \& Peterson 1965) in the 
quasar spectra.\footnote{Residual clumps (filaments) of neutral gas
remain and will produce the Lyman-alpha forest.}
We can use the observational limits on the Gunn-Peterson optical depth
as a test for cosmological models. In order to compute the Gunn-Peterson
optical depth, we generate random lines of sight through our simulation
box\footnote{Those lines of sight are taken to be random with respect to
the box orientation,
since at lower redshifts, when the nonlinear scale approaches the box
size, there appear purely artificial correlations between the density
distribution and the orientation of the computational box.}, and compute
the Lyman-alpha absorption $\tau(\lambda)$ along each line of sight.
We then compute the average opacity
of the intergalactic medium,
\begin{equation}
	\exp(-\bar\tau)\equiv{1\over N_{\rm ray}} \sum_{j=1}^{N_{\rm ray}}
	{1\over \lambda^{(j)}_2-\lambda^{(j)}_1}
	\int_{\lambda^{(j)}_1}^{\lambda^{(j)}_2} \exp\left(-\tau(\lambda)
	\right)d\lambda,
	\label{tauavg}
\end{equation}
where $N_{\rm ray}\sim1000$ is the number of lines of sight, and
$\lambda^{(j)}_1$ and $\lambda^{(j)}_2$ are wavelength limits for the
$j$ line of sight. This computation can be compared with observations
by Jenkins \& Ostriker\markcite{JO91} (1991), when expressed as
$1-D_A\equiv\exp(-\bar\tau)$. 

\placefig{
\epsscale{0.7}
\begin{figure}
\insertfigure{\figdir/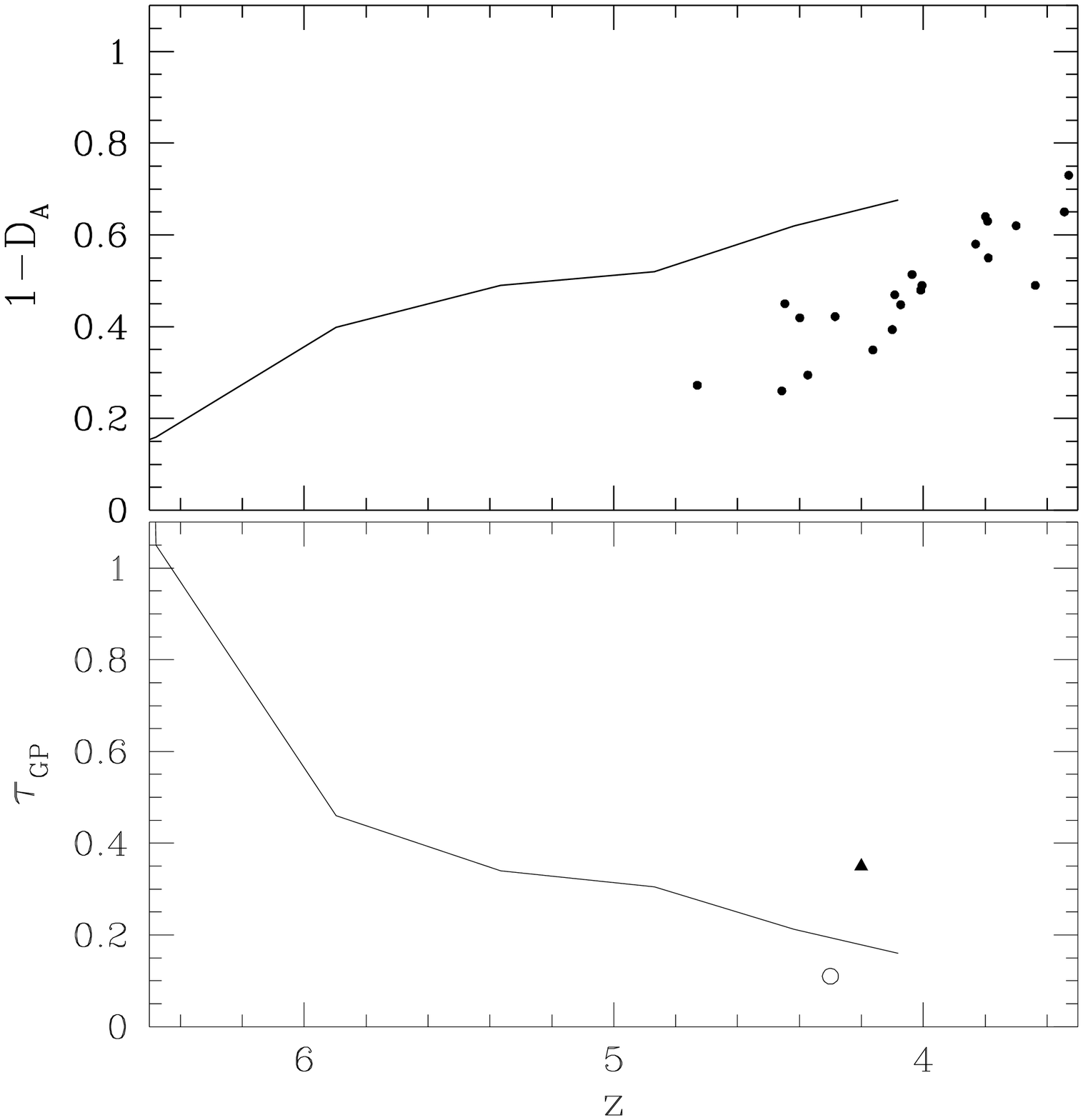}
\caption{\label{figDA}\capDA}
\end{figure}
}
We show in Figure \ref{figDA} (the upper panel) the data
from Jenkins \& Ostriker\markcite{JO91} (1991) as filled circles together
with our computation (the solid line). We note that our simulation predicts
somewhat too little absorption compared to the Jenkins \& Ostriker data,
but the general behavior is quite similar to the observations. We 
reiterate here that we do not expect to get an exact match with all existing
observational data on high redshift gas contents of the universe since
we simulate only one cosmological model which may not be correct, and
our simulations are subject to uncertainties due to the phenomenological
description of star formation.

We can also express the average absorption along the line of sight as
the the average Gunn-Peterson optical depth.
However, we must exclude the Lyman-alpha forest from the total
absorption since it is not counted for toward the total Gunn-Peterson optical
depth (Giallongo et al.\markcite{Gea94} 1994). 
We then compute the average Gunn-Peterson absorption
$\tau_{\rm GP}$
as the amount of absorption in the parts of the spectrum that lie above
the average opacity,
\begin{equation}
	\exp(-\tau_{\rm GP})\equiv{1\over N_{\rm ray}} \sum_{j=1}^{N_{\rm ray}}
	{\displaystyle\int_{\lambda^{(j)}_1}^{\lambda^{(j)}_2} 
	\exp\left(-\tau(\lambda)\right)\theta\left(\bar\tau-
	\tau(\lambda)\right)d\lambda\over\displaystyle
	\int_{\lambda^{(j)}_1}^{\lambda^{(j)}_2} \theta\left(\bar\tau-
	\tau(\lambda)\right)d\lambda},
	\label{taugp}
\end{equation}
where $\theta(x)$ is a step function.\footnote{This method of computing the
Gunn-Peterson optical depth is similar to the Miralda-Escude et 
al.\markcite{MCOR95} (1995) method for identifying Lyman-alpha lines.}

We show in the lower panel of Figure \ref{figDA} 
the Gunn-Peterson optical depth as a function
of redshift from our Run B. We also show observational upper 
limits from Jenkins \& Ostriker\markcite{JO91} (1991) as a solid triangle 
and from Giallongo et al.\markcite{Gea94} (1994) as an open circle. We again 
note that we predict too little absorption compared to the Jenkins \& Ostriker
data but too much absorption compared to the Giallongo et al.\ data.
Since
the Giallongo et al.\markcite{Gea94} (1994) limit is the most stringent one,
we adopt it for our analysis. We note that the result of our simulations,
while broadly consistent with observations, 
goes above the observational upper limit. This implies that the real
universe is more ionized than predicted by our simulations.
The final value for
$J_{21}$ is 0.34 at $z=4$, which is slightly below the currently favored
observational value as we have noted above. The more realistic value of
around $0.7-1.0$ would decrease the neutral hydrogen fraction and give
lower Gunn-Peterson optical depth. However, we cannot rescale the ionizing
intensity a posteriori preserving the accuracy of our simulations,
since the radiation field is tied up in the complex 
nonlinear evolution together with the intergalactic medium\footnote{In 
particular, even if most of the low density intergalactic medium is in
the ionization equilibrium with the radiation field, the value of the
temperature at every fluid element depends on the previous evolution
of the density and the radiation field in this fluid element
(Miralda-Escude \& Rees\markcite{MR94} 1994); since we
explicitly allow for the nonuniform radiation background, this dependence
cannot be parametrized in a simple form.}, and we would have to rerun
the whole simulation with new values for the radiative efficiencies for
our stellar particles as explained in \S 2.1. 

Given both the observational and theoretical uncertainties, the agreement
between the predicted and observed opacity of the intergalactic medium seems
quite adequate. It will be interesting to see if this rough agreement
survives when other cosmological models are considered.

\subsection{Metal Enrichment of the Intergalactic Medium}

Since in our simulations we follow the star formation with the phenomenological
approach, we also account for the dynamics of gas that was processed in
stars, and, is, therefore, enriched with heavy elements ({\it metals\/}).
However, since the IMF at high redshift is not known and may deviate
significantly from the standard IMF, we have no way to assign the value for
the metallicity in the processed gas reliably, but we still can study the
distribution of this metal enriched gas as compared to the distribution of
the primeval gas.

\placefig{
\begin{figure}
\insertfigure{\figdir/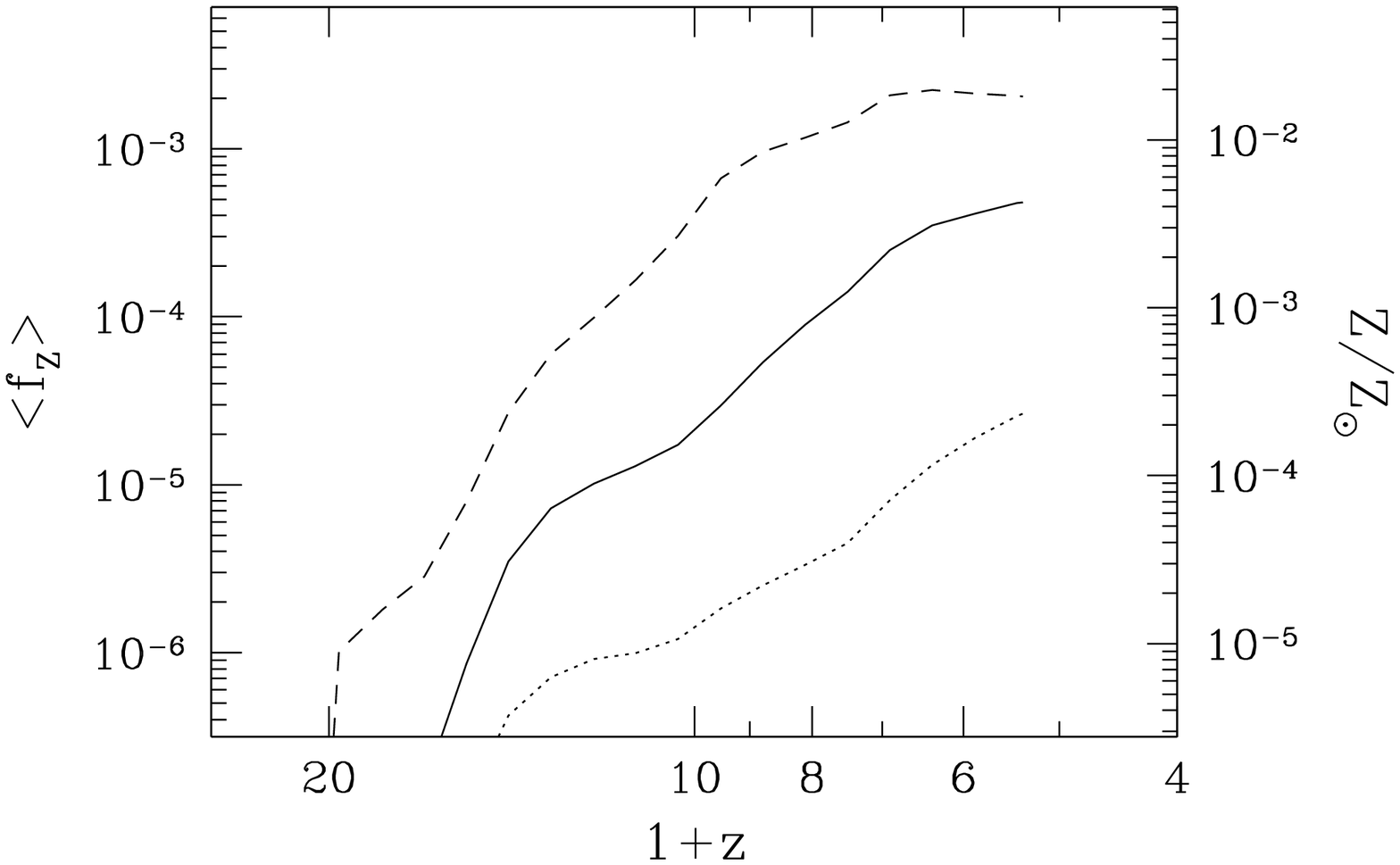}
\caption{\label{figFZ}\capFZ}
\end{figure}
}
In Figure \ref{figFZ} we plot the fraction $f_Z$ of the metal enriched gas 
in the total intergalactic gas as a function of redshift for our fiducial 
run A. Three different values correspond to different weights:
the mass weighted value is shown with the solid line, the volume weighted
value is shown with the dotted line, and the dashed line corresponds
to the weighting with the $\rho^{5/3}$ weight, which roughly reflects
the weighting by the column density of a Lyman-alpha absorber.
Here $f_Z$ is the ratio of the mass of the gas that was processed
in stars\footnote{We assume here that in average stars loose 10\% of their
total mass into the intergalactic medium; this amount does not include the
mass loss that is confined in the interstellar medium. If this amount is
different from 10\%, the plotted $f_Z$ should be adjusted simply in proportion
to the total fraction of stellar mass lost to the intergalactic medium.}
to the total mass of the intergalactic gas. If the ``yield'' for metal 
production with standard definition is $y$, than the ``metallicity'', or
a fraction of mass in elements heavier than helium is $9yf_Z$, given
our definitions. Thus, for $y=0.020$, which is characteristic for
protoellipticals as measured by the metallicity of the intracluster gas
(c.f.\ Silk\markcite{S96} 1996),
the final value of $\langle f_Z\rangle\approx5\times10^{-4}$ corresponds to
$\bar Z\approx 9\times10^{-5}$ or $\bar Z/Z_{\sun}\approx4.3\times10^{-3}$.
For a more standard value for the yield ($y=0.005$) we 
would obtain the final
metallicity of the order of $\bar Z/Z_{\sun}\approx1.0\times10^{-3}$.
We also note that by $z=4$ the 
volume average fraction of the metal enriched gas may reach a value of
about 0.05\%.

\placefig{
\epsscale{0.65}
\begin{figure}
\insertfigure{\figdir/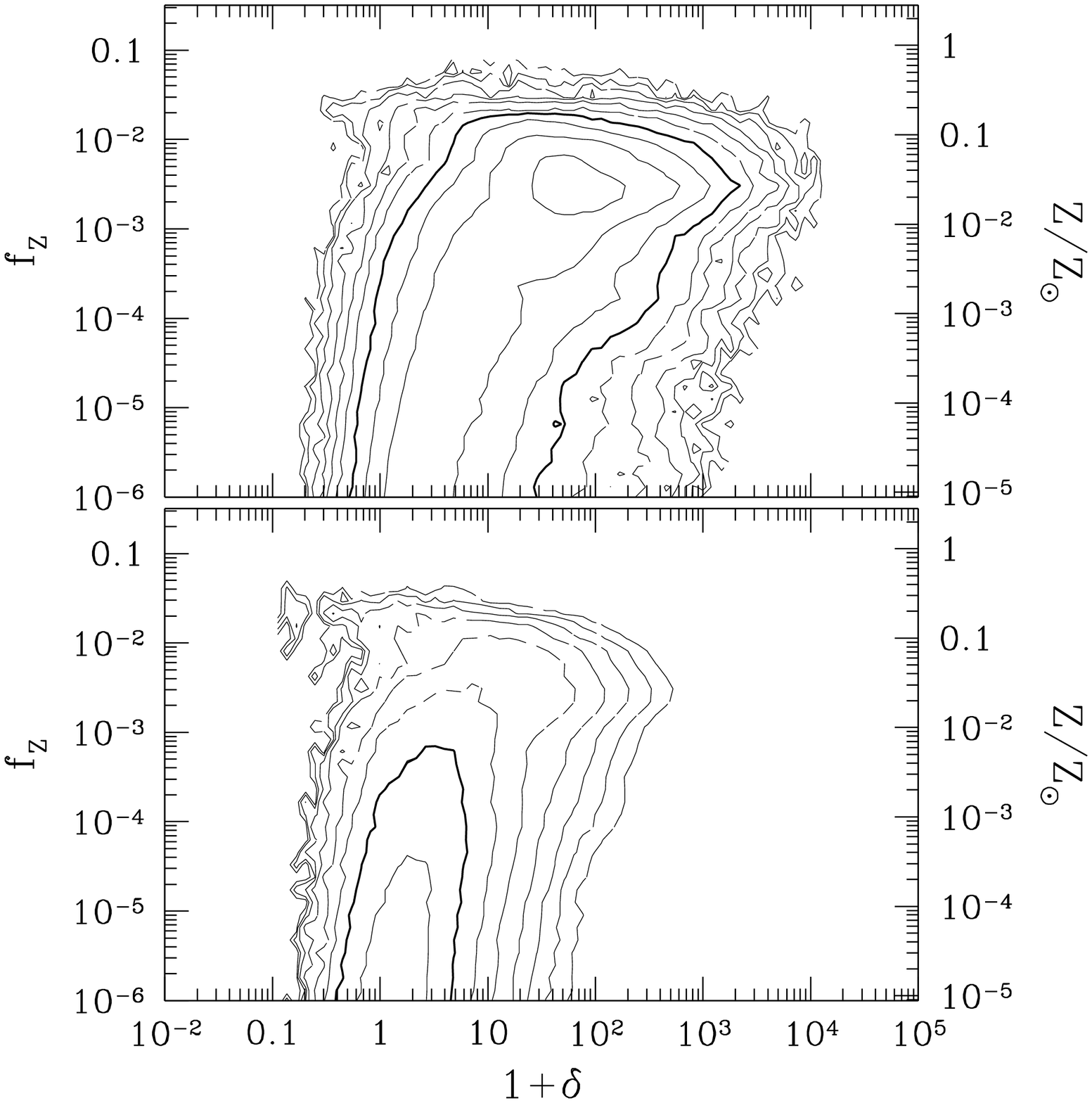}
\caption{\label{figMF}\capMF}
\end{figure}
}
While Fig.\ \ref{figFZ} shows a rough consistency with observations, it
does not show us where this metal enriched gas is
actually located and it understates the variability in heavy element
abundances at a given epoch. 
In Figure \ref{figMF} we show the probability distribution
to find a fluid element with the particular values of the fraction of the
metal enriched gas and the total density of the intergalactic gas, as 
measured by the overdensity $\delta\equiv\rho/\bar\rho-1$. The upper panel 
shows the mass weighted, and the lower panel shows the volume weighted
distributions. We note that most of the volume has $f_Z$ well below 0.1\%,
whereas a considerable fraction of mass has $f_Z$ of the order of
$0.1-1.0$\%. We also note that at low overdensities, which are characteristic
of the low column density Lyman-alpha forest (c.f.\ Miralda-Escude 
et al.\markcite{MCOR96} 1996; Hui, Gnedin, \& Zhang\markcite{HGZ96} 1996), 
the range of existing metallicities is extremely broad, while 
the average value is quite low. This is in a qualitative agreement with
the conclusion by Rauch, Haehnelt, \& Steinmetz\markcite{RHS96} (1996)
who found that observed scatter in the abundance ratios of the QSO
metal absorption systems exceeds that predicted by simulations with
the uniform metallicity.

\placefig{
\begin{figure}
\insertfigure{\figdir/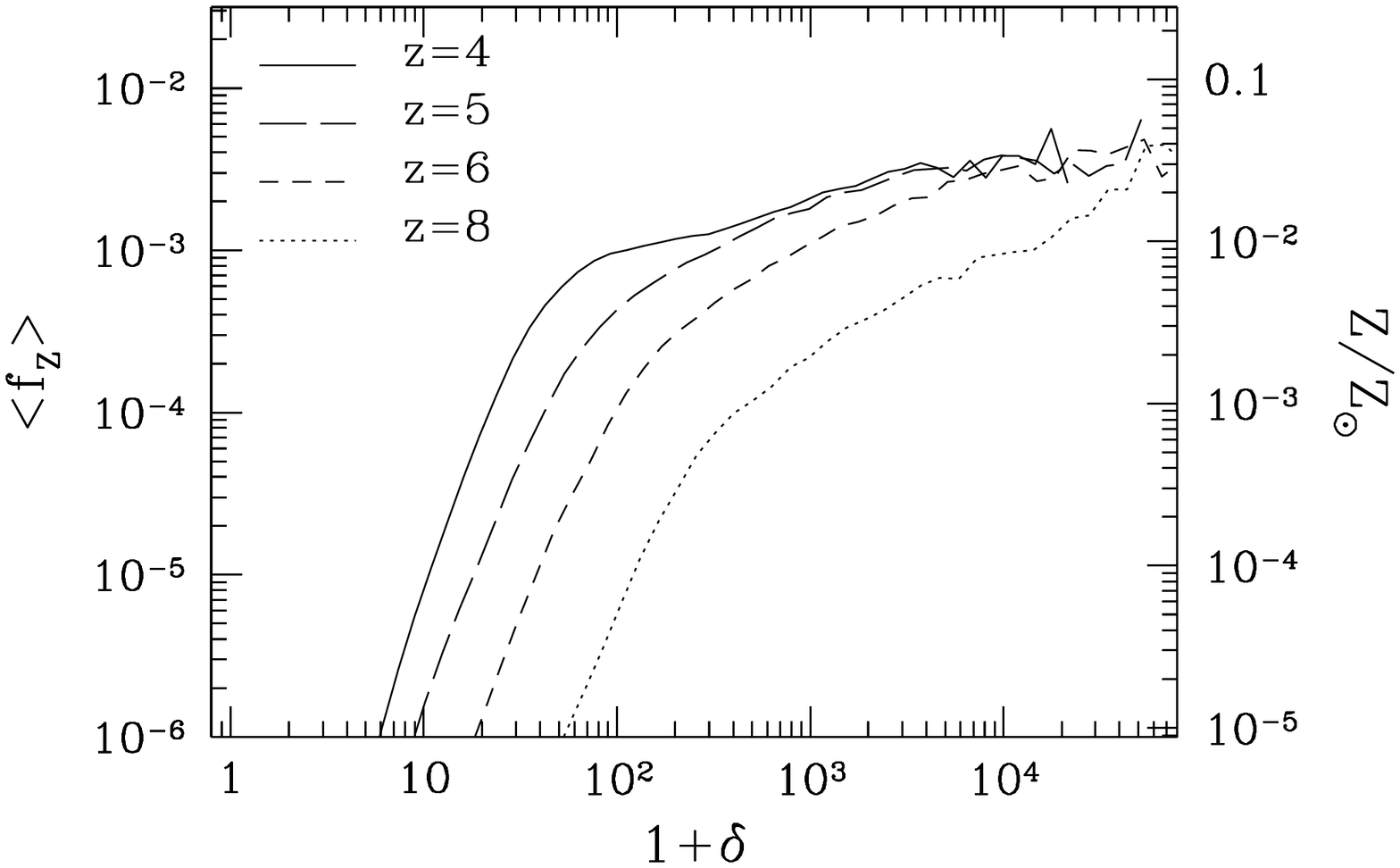}
\caption{\label{figZD}\capZD}
\end{figure}
}
To quantify this effect further, we plot in Figure \ref{figZD} the average
fraction of the metal enriched gas as a function of the total gas density
at four different redshifts.
In other words, Fig.\ \ref{figZD} is a Fig.\ \ref{figMF} collapsed along the
$y$-direction. 
It is interesting to note that the final values of the metallicity in the
high density regions from which galaxies will be made approach 
$Z/Z_{\sun}\approx1/30$, which is close to the minimum metallicity found in
the oldest and most metal poor Population II component of our
Galaxy. More that that, this value does not change significantly with time
after $z\sim8$. We also point out that there seems to be steady diffusion
of heavy elements into low density regions.

It is remarkable that the average metallicity is only a
weak function for wide range of overdensities above $\delta\sim10^2$, and
it decreases very fast as the overdensities become smaller. This effect
allows us to make a specific prediction that the metallicity in the 
Lyman-alpha forest should decrease sharply for the column densities below about
$10^{13.5-14.5}\dim{cm}^{-2}$ as measured at $z=3$ (which roughly corresponds
to $\delta\sim30$, see Hui et al.\markcite{HGZ96} 1996). 
Here we allow ourselves a 
broad error bar in the quoted value of the characteristic column density
to account for the possible range of cosmological models and uncertainties
in the simulations, both numerical uncertainties (like the limitations of
the phenomenological star formation algorithm), and uncertainties in the
input physics (like our lack of knowledge of the IMF at high redshift or the
precise value of the cosmic baryon density).
Recent measurements of metal abundances in the Lyman-alpha forest 
(Songalia \& Cowie\markcite{SC96} 1996) 
found heavy elements in a large fraction ($\ga75\%$
of all Lyman-alpha systems with column densities in excess of $10^{14.5}
\dim{cm}^{-2}$). Our prediction then implies that, when those studies are 
extended to lower column densities, the sharp decline in the fraction of
Lyman-alpha systems with measurable heavy elements abundances will be found
at column densities of about $10^{13.5-14.0}\dim{cm}^{-2}$. The specific value
of the column density at which the metal abundance of the Lyman-alpha forest 
declines can serve as a new test for cosmological models.

\placefig{
\begin{figure}
\epsscale{1.0}
\insertfigure{\figdir/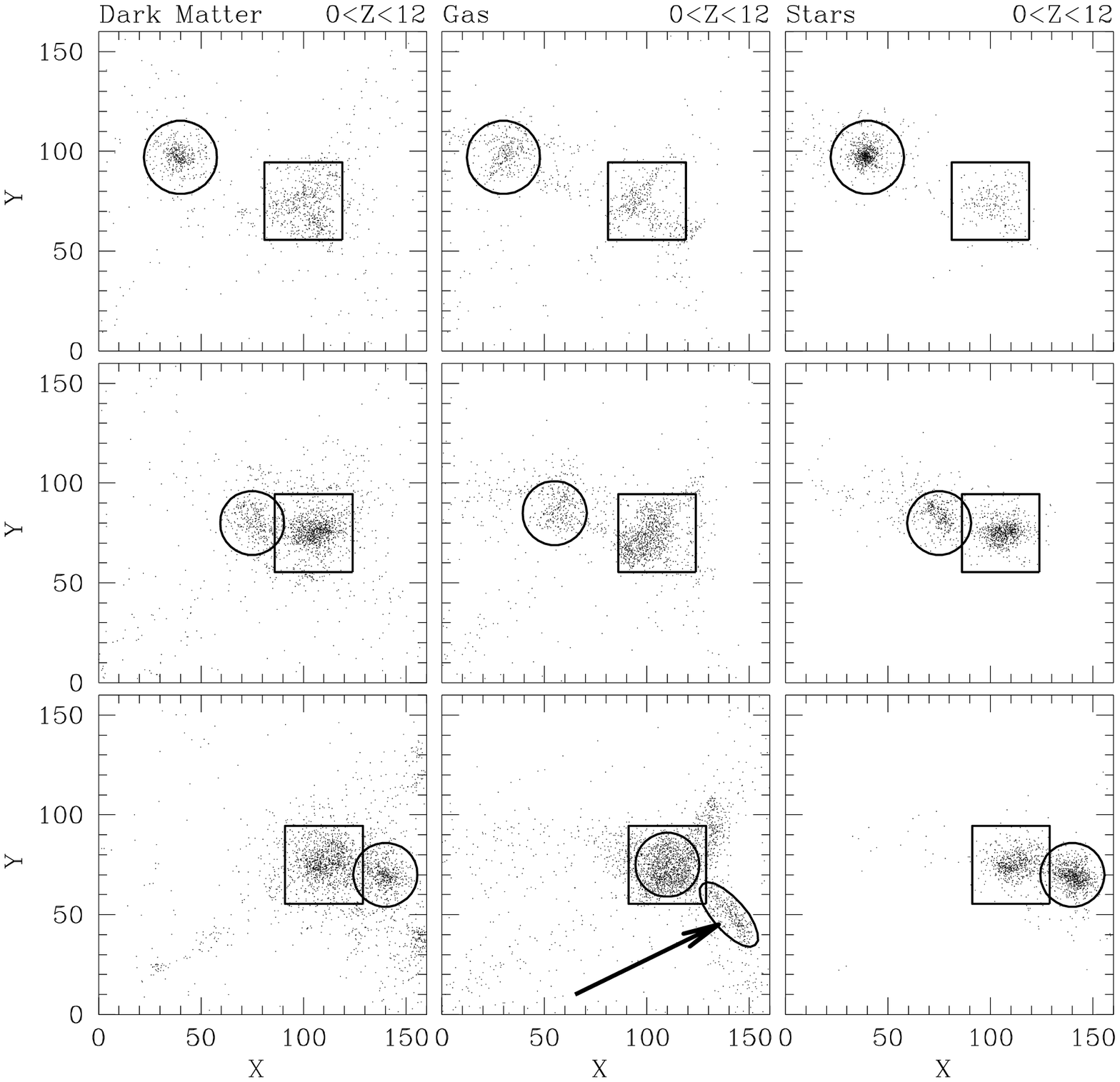}
\caption{\label{figZE}\capZE}
\end{figure}
}
The heavy elements are produced in the regions of star formation, where the
gas density is high and the potential well is deep, and, therefore one can
expect that they are effectively trapped inside the (proto-)galaxies, 
contrary to our finding that they are widely distributed even in the
low density regions. We must therefore address the question of 
how heavy elements
eventually get into the low density regions. Since our simulations do not
include any special diffusion processes that may be responsible for 
distributing the metal enriched gas over a large range of densities, we
conclude that there must exist a mechanism which works entirely within the
framework of hierarchical clustering. Figure \ref{figZE} demonstrates how
this may have happened. We show the distribution of the dark matter 
(the left column), the cosmic gas (the middle column), and the stars (the right
column) at three different redshifts: $z=5.4$ (the upper row), $z=4.9$
(the middle row), and $z=4.3$ (the lower row) represented by 
particles\footnote{While dark matter and stars are numerically represented
as particles in the SLH approach, the gas component is numerically
represented by a distorted quasi-Lagrangian mesh; we convert 
(for display only) cells of this
mesh into gas particles for the sake of simplicity and uniform presentation.}
in a thin slice with the width of $12h^{-1}$ comoving 
kpc.\footnote{While dark matter particles all have the same mass, the gas
particles and the stellar particles have different masses and the visual
appearance of the particle density does not necessarily correspond to the
actual gas or stellar density in the slice.}
All distances in the slice are in $h^{-1}$
comoving kpc. There are two separate objects at $z=5.4$ marked with the
square and the circle that are undergoing merger at lower 
redshift.\footnote{The objects may have different appearance at 
different redshifts
due to the thinness of the slice: particles comprising the objects may
enter and leave the slice at different redshifts; we show only a subset of all
particles in the each object for the sake of clearness.}
At the
lower row at $z=4.3$ the dark matter and the star components of the smaller
object (circle) passed close to the center of the larger object, while the
gas components of both objects collided in a gigantic shock and merged
at the first impact; as the result the stream of high density gas was
ejected from the merged object (marked with the oval in the lower middle
panel, and pointed to by the arrow). The gas ejected came mostly from 
the larger object, and was heavily enriched by heavy elements. In the next
Hubble time or so it will be dispersed and mixed with the 
surrounding primeval intergalactic gas, producing high metal abundance
even in low density regions lying close to the merged object. But the
low density gas far from any merger would still be uncontaminated. 
Thus, the metal enrichment of the intergalactic gas will be highly 
inhomogeneous, as shown in Fig.\ \ref{figFZ}. 

We also note that the fact that we demonstrated a 
working mechanism for the heavy 
element enrichment does not of course exclude the possibility that there
exist other mechanisms which distribute metals in the low
density intergalactic gas. In particular, supernova explosions, 
which are included in the work of Cen \& Ostriker\markcite{CO92} (1992) ,
are effective at dispersing metal enriched gas.

\subsection{The 21cm Emission from High Redshift}

The neutral gas at high redshift is surrounded by the thermal bath of CMB
photons, which has a temperature much higher than the excitation temperature of
the $21\dim{cm}$ line. The hydrogen atoms, being excited by the CMB photons,
spontaneously decay emitting a $21\dim{cm}$ line radiation, which being
constantly redshifted during the evolution of the universe, produces 
continuum signal at the telescope. As the universe evolves and structure
develops, velocity focusing in the converging flows will produce enhanced
emission at lower redshifts until reionization, when the neutral hydrogen
fraction drops by several orders of magnitude, and $21\dim{cm}$ emission
declines sharply at the frequency corresponding the $21\dim{cm}$ redshifted
to the epoch of reionization. 

\placefig{
\begin{figure}
\epsscale{0.7}
\insertfigure{\figdir/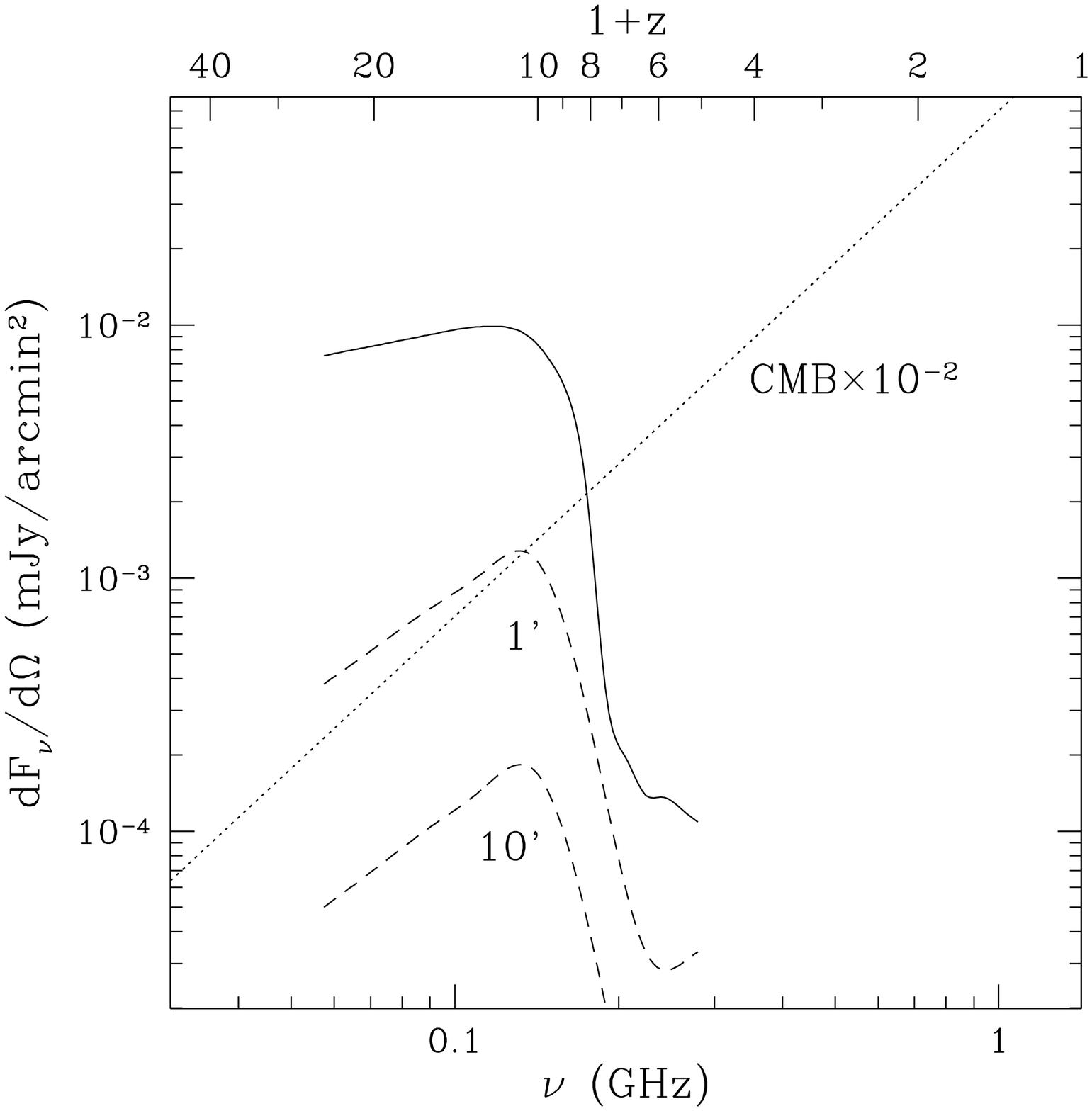}
\caption{\label{figEM}\capEM}
\end{figure}
}
We show in Figure \ref{figEM} the spectrum of the redshifted $21\dim{cm}$
line as would be observed on Earth (the solid line), 
together with the contribution from
the Cosmic Microwave background reduced by a factor of a hundred to fit
on the same scale (the dotted line). 
The very different spectral signature of the $21\dim{cm}$
radiation makes it possible, at least in principle, to distinguish it from
the CMB. However, the amplitude of the signal is still 
substantially lower than the sensitivity of all existing or being planned
radio telescopes (including the Square Kilometer Array).
Nevertheless, when (and if) this measurement is performed, the reward will be
quite substantial: since reionization is so sudden, and, therefore, the
break in the spectrum at the frequency $\nu_b=\nu_0/(1+z_{\rm reion})$ is
also very sharp, the measurement of the break frequency will give a very
accurate value for the redshift of reionization, which, in turn, can place
a strong constraint on both cosmological models and physical parameters at
high redshift. 

We also show with the dashed line the rms fluctuations in the $21\dim{cm}$
radiation on 1 and 10 arcmin angular scale. Since the 1 arcmin corresponds
to the comoving scale of around $1.7h^{-1}\dim{Mpc}$ at $z=7$ in the 
adopted cosmological model, the density fluctuations on this scale are
of the order of 20\%; in the sky those fluctuations will result in 
the fluctuations in the redshifted $21\dim{cm}$ radiation. Even if the
isotropic component of the $21\dim{cm}$ radiation is only 1\% of the
CMB fluctuations, because of the incredible smoothness of the CMB sky,
$\Delta T/T\sim10^{-5}$, fluctuations in the $21\dim{cm}$ will dominate
the CMB fluctuations by a factor of $10^{5}\times10^{-2}\times0.2=200$!
Since at lower frequencies (earlier times) the density fluctuations
are smaller, there is a peak in the spectrum of rms fluctuations in the
$21\dim{cm}$ radiation close to the frequency corresponding to the
$21\dim{cm}$ frequency redshifted to $z_{\rm rei}$. This would provide
an alternative (and, likely, much more efficient) way to measure the redshift
of reionization from the $21\dim{cm}$ from high redshift gas.

\placefig{
\begin{figure}
\insertfigure{\figdir/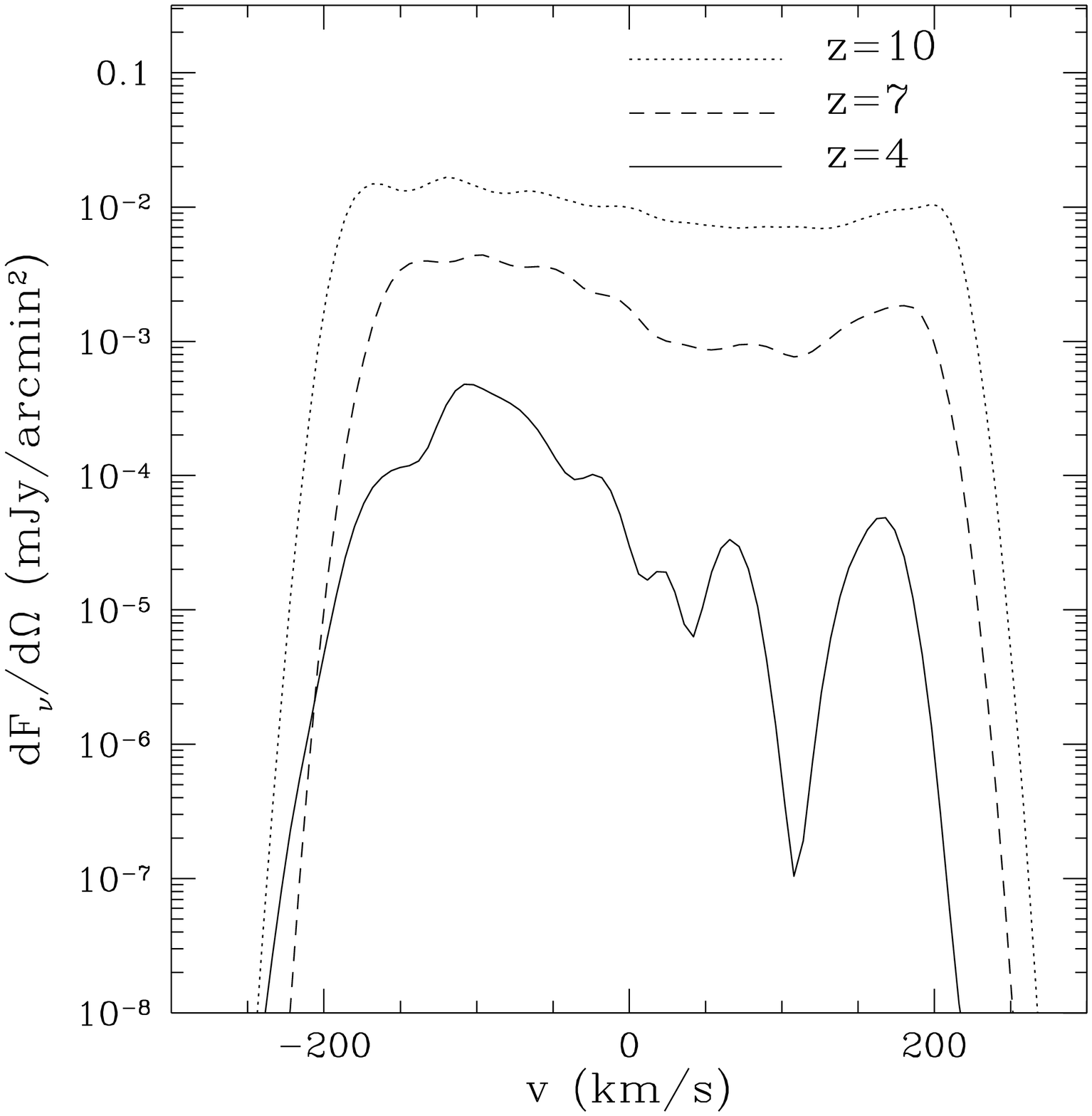}
\caption{\label{figEF}\capEF}
\end{figure}
}
The spectrum in Fig.\ \ref{figEM} shows the $21\dim{cm}$ emission as measured
by a broad filter with $200\dim{km}/\dim{s}$ width. However, since after
reionization the most of neutral hydrogen is in optically thick lumps 
(protogalaxies), the final structure of the $21\dim{cm}$ emission at
lower redshift would consist of a combination of several lines rather than
of a continuum emission. Figure \ref{figEF} serves to illustrate this
conclusion. We show the $21\dim{cm}$ emission as a function of frequency
as measured in velocity units at three different redshifts each centered 
at the frequency corresponding to the redshifted $21\dim{cm}$ line. The
emission terminates at velocities larger than $200\dim{km}/\dim{s}$
due to the final size of the simulation box. We note that while at high
redshift emission is almost uniform in frequency, at lower redshift, after
reionization, it consists of several overlapping lines.

\subsection{Evolution of the Background Radiation}

\placefig{
\begin{figure}
\insertfigure{\figdir/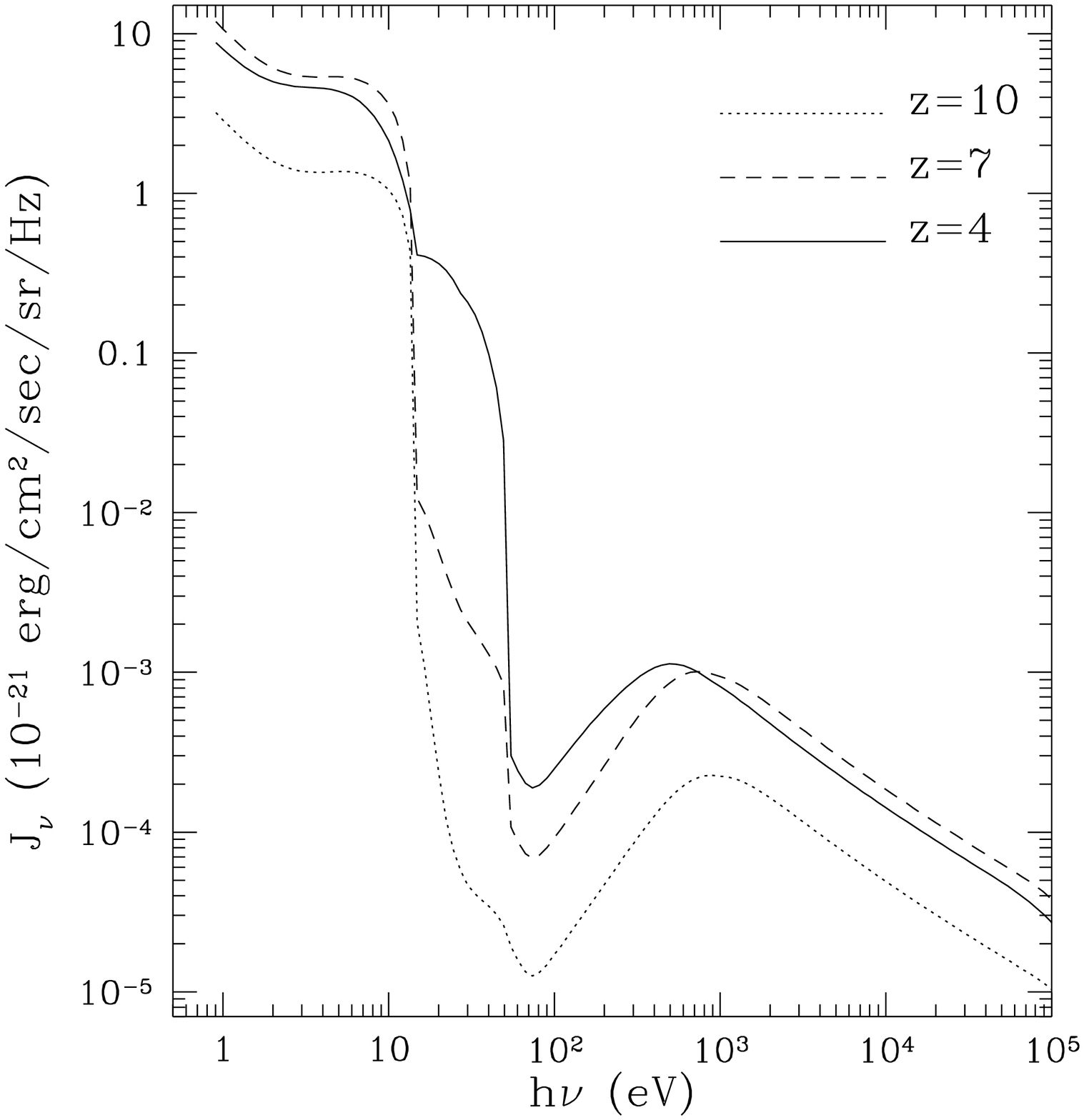}
\caption{\label{figSS}\capSS}
\end{figure}
}
The radiation emitted by stellar- and quasar-like sources participates in a
complex interaction with the highly inhomogeneous intergalactic medium. The
interplay between ionization and recombination processes in the intergalactic
gas, coupled with star formation and development of cosmic structure
determines the shape and the amplitude of the background radiation which
are not guaranteed to be monotonic in frequency or time. As an illustration,
we show in Figure \ref{figSS} the spectra of the background radiation at three
different redshifts. This is qualitatively very similar to figure 8 of
Ostriker \& Cen\markcite{OC96} (1996) despite the considerably more 
detailed treatment of radiative processes in this paper.
We note that in the $13.6\dim{eV}<h\nu\lesssim1\dim{keV}$ range
the background radiation spectrum has a complex shape with pronounced
ionization edges at $13.6\dim{eV}$, $24.6\dim{eV}$, and $54.4\dim{eV}$.
The neutral helium ionization edge at $24.6\dim{eV}$ becomes weak after
helium becomes at least singly ionized somewhat prior to hydrogen 
reionization at $z=7$, but the hydrogen ionization edge at $13.6\dim{eV}$
is still easily observed even at $z=4$, well after reionization of 
the universe. Examination of Fig.\ \ref{figSS} shows that the assumption
of a power-law (or other simple form) for the background radiation field,
which is an approximation adopted by several authors, would have been 
extremely inaccurate due to the strong effects of atomic ionization edges.

\subsection{Properties of Proto-galaxies}

We now concentrate on general properties of objects formed in our simulations, 
which we identify with small proto-galaxies. Paper I addressed in some
detail the question
of the various stellar Populations and star formation rates
at different epochs. Here we
will discuss some of the internal properties of those objects. 

\placefig{
\begin{figure}
\insertfigure{\figdir/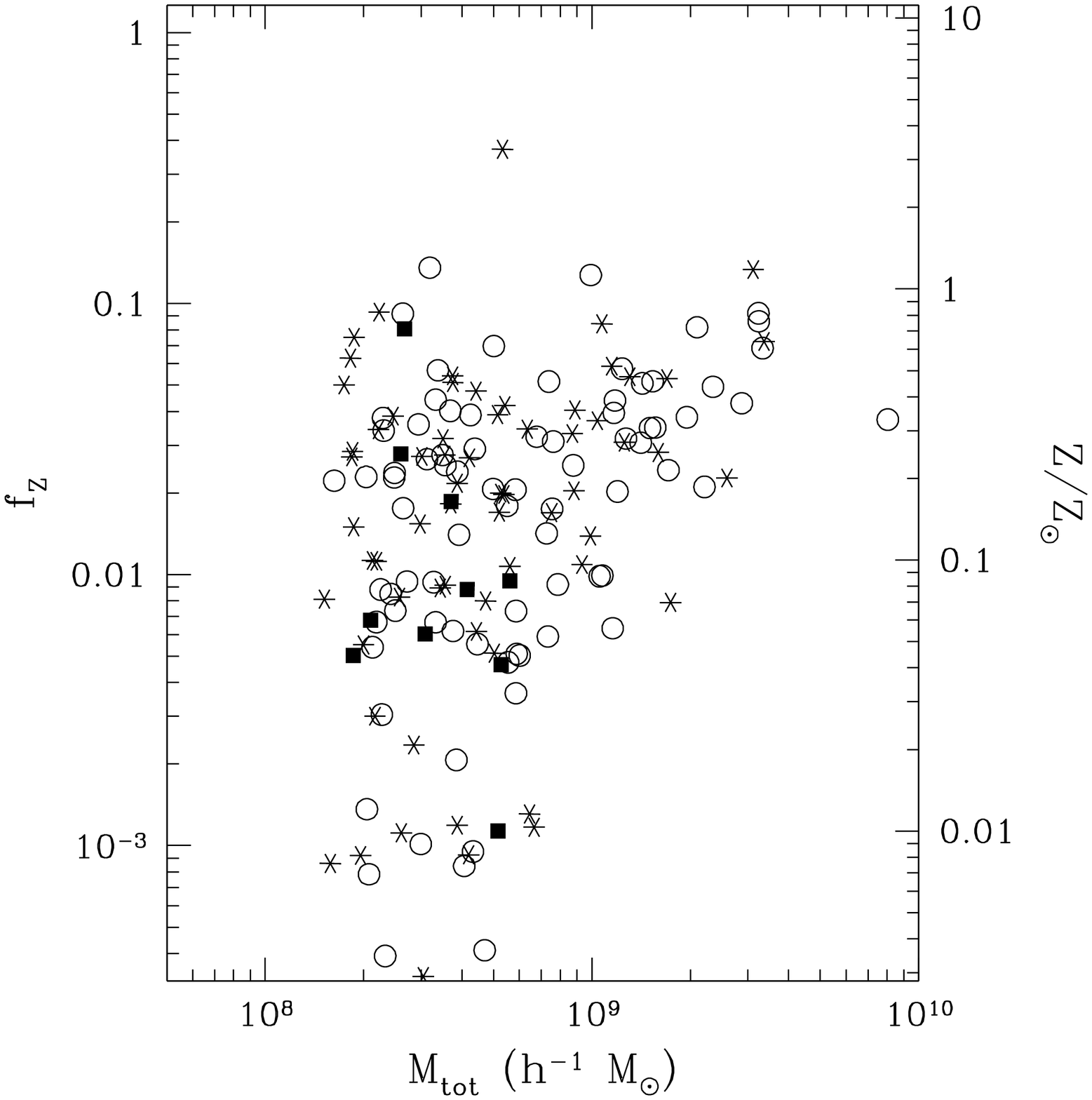}
\caption{\label{figZM}\capZM}
\end{figure}
}
Figure \ref{figZM} shows the fraction of metal enriched gas as a function
of total mass for all objects with the total mass $M_{\rm tot}>10^8h^{-1}
M_{\sun}$ and stellar mass $M_*>10^5h^{-1}M_{\sun}$ at three different 
redshifts as marked with different symbols: $z=9.3$ (solid squares),
$z=5.9$ (stars), and $z=4.3$ (open circles). We note that there exists a
large fraction of objects with as much as 10\% of metal enriched gas, which
corresponds to solar metallicity if the high value of the yield
$y=0.020$ is adopted as discussed above. We also point out existence of
an object at $z\approx6$ with at least the solar metallicity (if the standard
value for the yield $y=0.005$ is adopted) or several times the solar 
metallicity if the higher value of yield is accepted. We therefore conclude
that hierarchical clustering models have no difficulty in explaining
high metallicity observed in quasars at high redshifts. 
We emphasize here that due
to the small size of our computational box, our simulations are strongly
biased {\it against\/} finding very bright massive quasars as those are
expected to be rare events in the gaussian initial conditions.

\placefig{
\begin{figure}
\inserttwofigures{\figdir/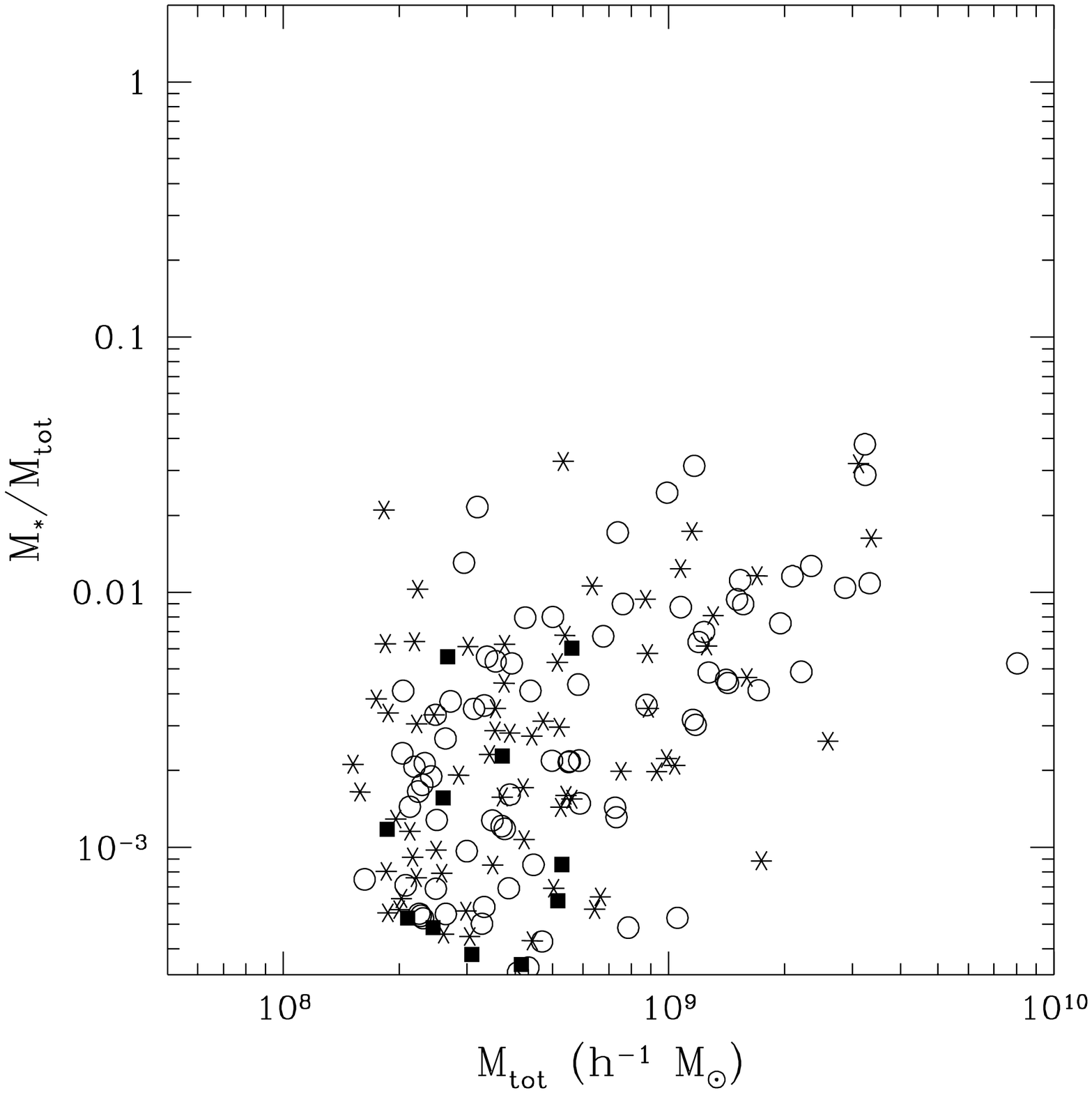}{\figdir/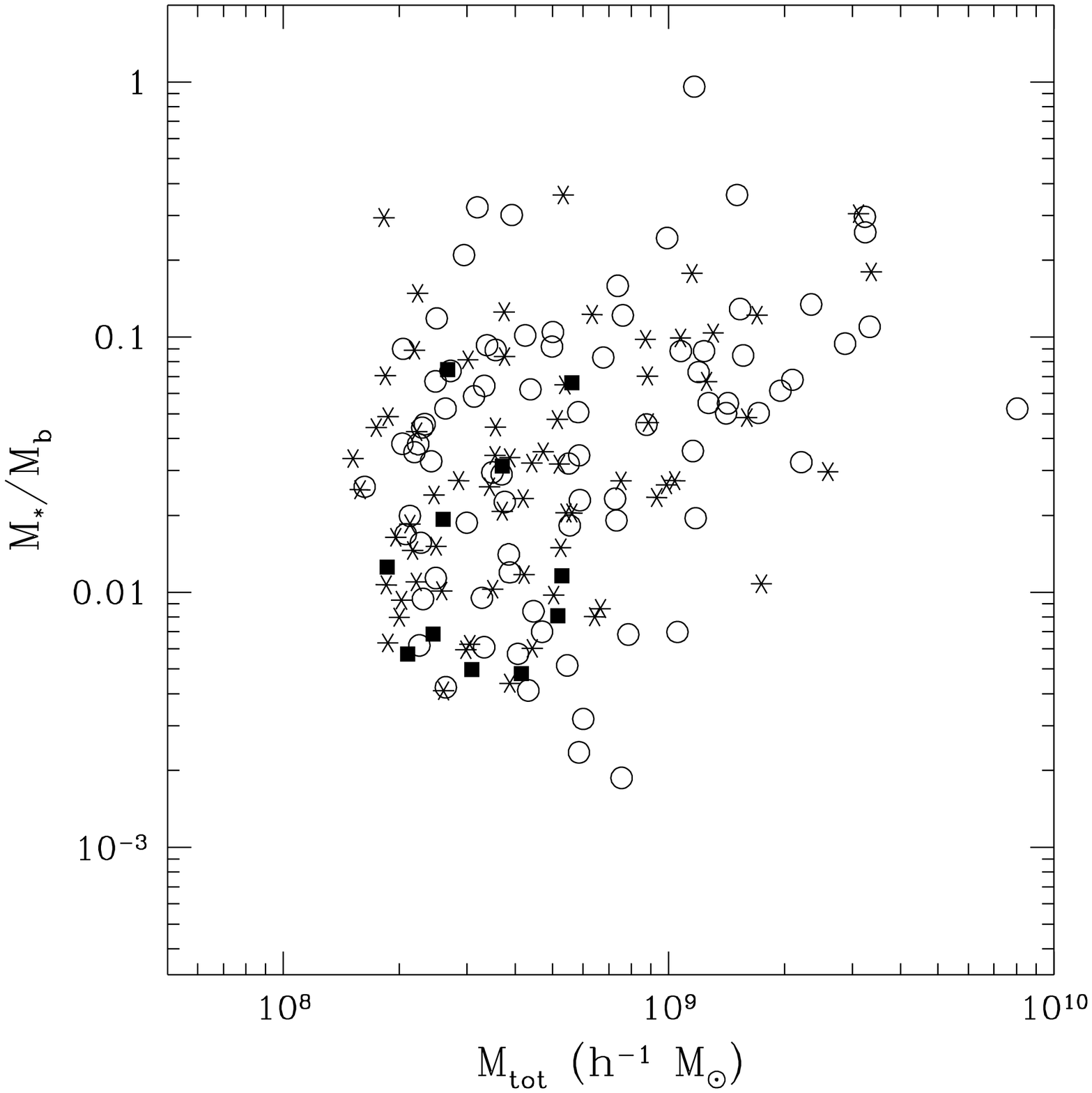}
\caption{\label{figFM}\capFM}
\end{figure}
}
In Figure \ref{figFM} we show the mass in stars $M_*$ as a fraction of
the total mass $M_{\rm tot}$ ({\it a\/}) and the baryonic mass
$M_b$ ({\it b\/}) of a given object as a function fo the total mass of
the object at three different redshifts as described above. A considerable 
fraction of all objects have already turned almost a half of 
their gas into stars by $z\approx4$, however, even at the centers those objects
are still dominated by the dark matter; stars constitute on average one third
of the total density at the center by $z\approx4$ for objects with the 
total mass above $10^9h^{-1}M_{\sun}$. This is mostly due to our finite
resolution ($1h^{-1}$ comoving kpc), which corresponds to the physical scale
of around $300\dim{pc}$ at $z\approx4$, and partly due to the fact that
objects in Fig.\ \ref{figFM} are still efficiently forming stars and actively
accreting intergalactic gas which is cooling and concentrating at the
center of objects, thus increasing the role of baryons compared to the
dark matter with time, and, therefore, if we were able to continue our
simulations further in time with the appropriate allowance for the missing
large-scale waves, we would observe the increasing fraction of baryons
at the centers of the proto-galaxies.

\placefig{
\begin{figure}
\inserttwofigures{\figdir/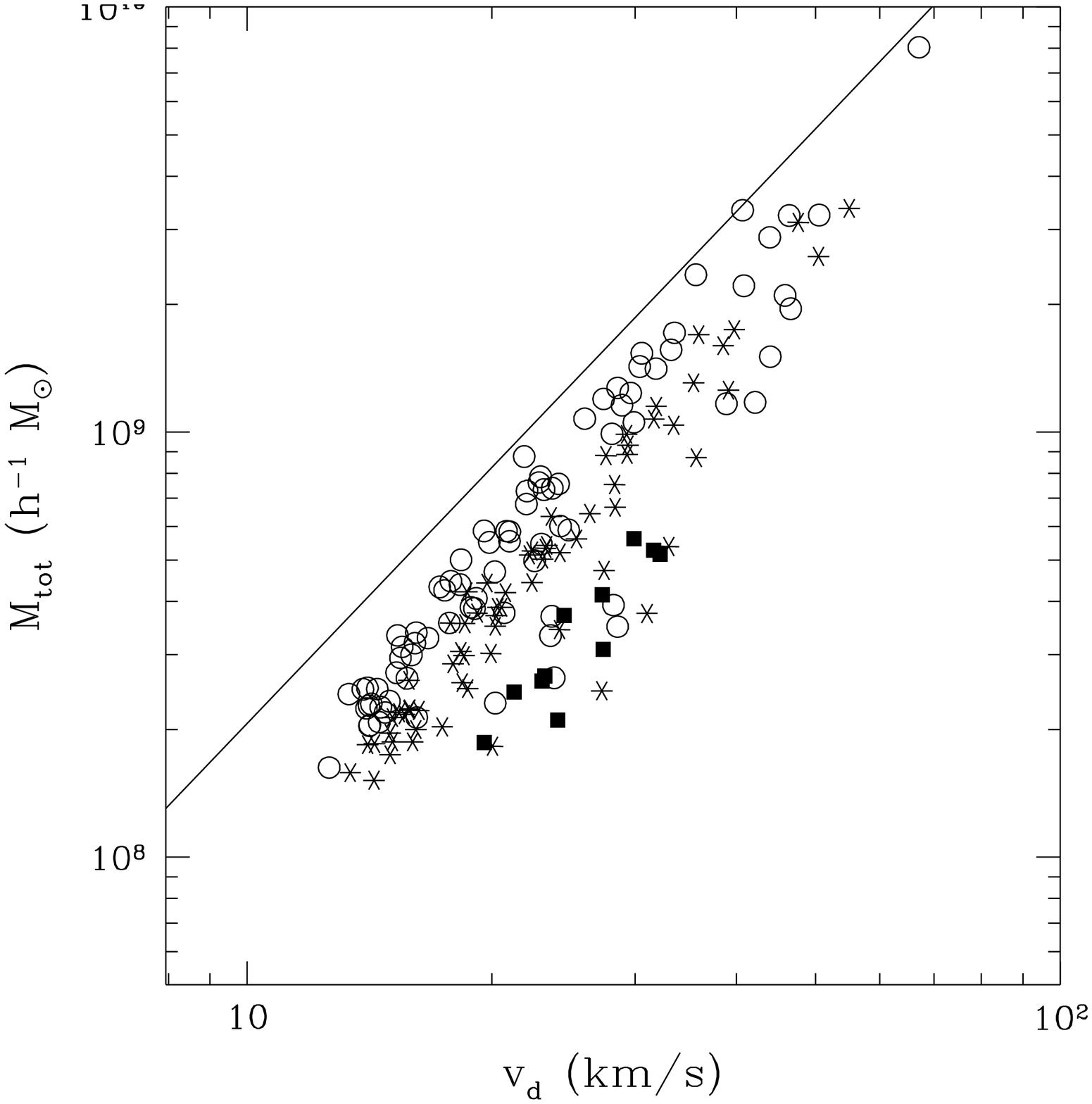}{\figdir/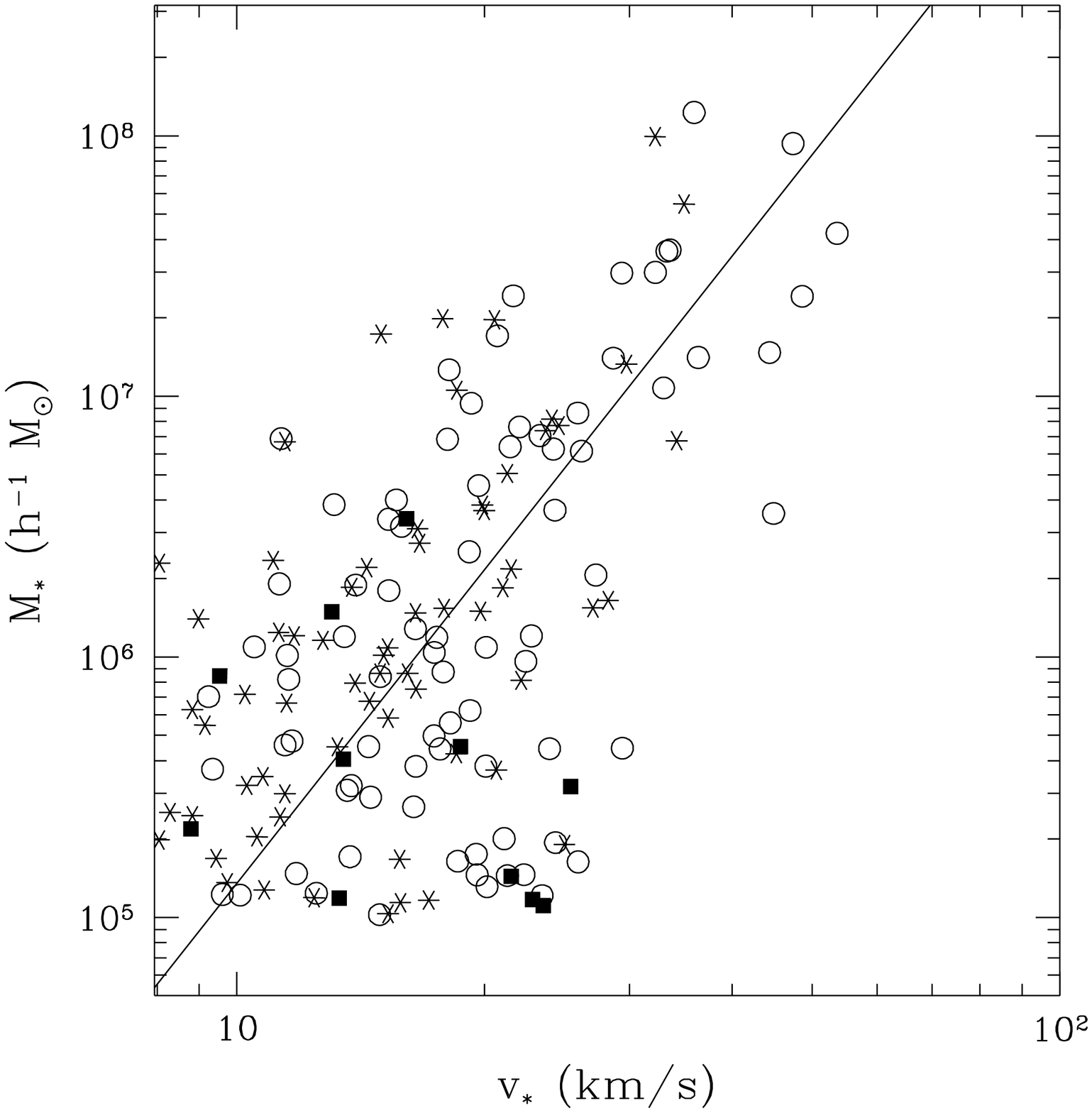}
\caption{\label{figMV}\capMV}
\end{figure}
}
Let us now discuss the gravitational structure of proto-galaxies. We show
in Figure \ref{figMV}a the total mass for all objects as a function of
dark matter velocity dispersion at three different redshifts as explained
above. The solid line shows the power-law dependence with the power-law index
of 2. The fact that our objects follow this power-law shape indicates
that they are close to the virial equilibrium. In Figure \ref{figMV}b we
plot the total stellar mass as the function of the stellar velocity dispersion,
and the solid line is the following law:
\begin{equation}
	M_{\rm tot} = 10^{10.5}h^{-1}M_{\sun}
	\left(v_*\over220\dim{km}/\dim{s}\right)^4,
	\label{tflaw}
\end{equation}
and the power-law index is now 4 instead of 2. If we compare this dependence
with the Tully-Fisher or Faber-Jackson relation we will get a good agreement
provided that the average stellar mass-to-light ratio is $\Upsilon=
10^{0.5}h=2.2$, which is a reasonable number consistent with the solar
neighborhood. We also point out that the
normalization of our ``Faber-Jackson'' relation is roughly 
independent of redshift.
The scatter is apparently substantially larger than the observed scatter
in large galaxies (about 1.5-1.7 magnitudes), 
but it mainly arises from the finite mass resolution of
our simulations: if the most massive objects in our simulations contain
many thousands of dark matter particle and even larger 
number of gas particles, they contain a much
smaller number of stellar particles.\footnote{The largest object in the 
simulation has 21,000 dark matter particles, 23,000 gas particles and 
2,200 stellar particles, whereas the objects at the bottom of Fig.\ \ref{figMV}b
have only 30-40 stellar particles.}
We also emphasize that the active merging is
going on around all massive objects in our simulation which tends to increase
the scatter.

\section{Conclusions}

We demonstrate by means of a high resolution numerical hydrodynamic simulation
how reheating and reionization of the universe occur in the CDM+$\Lambda$
cosmological model. We incorporate in our treatment essentially all 
relevant physical processes
which are important for assessing the evolution of average properties of the
universe, with one free parameter -- the efficiency of high mass star
formation -- determined by matching the computed and observed background at
redshift $z=4$.

We conclude that a generic prediction of models where the universe is
reheated and reionized by high energy radiation from stellar-like and
quasar-like sources is that slow reheating happens first followed by
sudden reionization. We also demonstrate that hierarchical clustering models
are broadly consistent (with the appropriate choice of parameters) with
the measured Gunn-Peterson opacity of the intergalactic medium. 

The average heavy element abundance in the intergalactic medium can
reach values of $1/200$ of solar by $z=4$ with the large dispersion
in the low density regions ($\delta\la30$). The metal abundances
in the high density regions are more uniform and can reach $1/30$ in average,
but there exist individual bound object (proto-galaxies) that can have
solar or higher abundance of heavy elements at $z\sim6$. This fact implies
that hierarchical clustering models like the one studied in this paper
can easily explain observed high metallicity of quasars at $z=5$.
We also propose a working mechanism based on mergers for transporting the
heavy elements, initially locked in the high density regions, into the
lower density intergalactic medium, where they are observed in 
(sufficiently high column density) Lyman-alpha absorbers. 

Reionization, which suddenly reduces the neutral hydrogen abundance by
several orders of magnitude, leaves a distinct signature in the redshifted
$21\dim{cm}$ emission from the intergalactic hydrogen. While this emission 
cannot be measured by existing or proposed meter-wavelength radio
telescopes, which lack sensitivity by a factor of several hundred, future
ultra-high sensitivity experiments may detect this emission, giving us
a very strong (due to the suddenness of reionization) constraint on 
the evolution of the universe at high redshift.

The Faber-Jackson relation, the correlation between the luminosity and
the velocity dispersion of a galaxy is reproduced by our simulations
(assuming constant mass-to-light ratio) with
correct amplitude and slope, indicating that this phenomenon can be
explained entirely based on hierarchical clustering model and simple
cooling arguments. Both, the slope and the amplitude of our Faber-Jackson
relation, are independent of redshift within numerical uncertainties.

\acknowledgements

The authors would like to express their gratitude to Prof.\ Martin Rees for
numerous fruitful discussions and valuable comments.
This work was supported by NSF grant AST-9318185
awarded to the Grand Challenge Cosmology Consortium.

\appendix

\section{Radiative Transfer Equations in the Expanding Universe}

Let us start with deriving the full radiative transfer equation in 
comoving coordinates. Let $F(t,x^i,p^k)$ be the distribution function for
photons in comoving coordinates $x^i$ and comoving momentum
$$
	p^k = a {h\nu\over c} n^k,
$$
where $a\equiv a(t)$ is the scale factor, $h$ is the Planck constant,
$\nu$ is the photon frequency, and $n^k$ is the unit vector in the
direction of photon propagation. By definition (Peebles 1980),
$$
	N_\gamma = \int F(t,x^i,p^k) d^3x\, d^3p
$$
is the number of photons in the universe; $N_\gamma$ changes only due to
photon absorption or emission in physical processes (ionization, 
recombination, cooling, etc). We can, therefore,
write down the continuity equation for the function $F$ in the 
phase-space $(x^i,p^k)$:
\begin{equation}
	{\partial F\over\partial t} + 
	{\partial\over\partial x^i}\left(\dot{x}^i F\right) +
	{\partial\over\partial p^k}\left(\dot{p}^k F\right) =
	{\rm RHS}(F),
	\label{Feq}
\end{equation}
where the term ${\rm RHS}(F)$ denotes photon absorption and emission in
physical processes.

It is customary, however, to use specific intensity $J_\nu(t,x^i,n^k)$
of radiation instead of $F(t,x^i,p^k)$, where, by definition,
$J_\nu d\nu\,d\Omega\,dA\,dt$ is the energy of photons with frequences
from $\nu$ to $\nu+d\nu$ passing through
the area $dA$ in the time interval $dt$ in the solid angle $d\Omega$
around $n^k$. It is easy to convert from $F$ to $J_\nu$ recalling
that $d^3x=c\,dt\,dA$ and $d^3p=p^2dp\,d\Omega$:
\begin{equation}
	J_\nu = h\nu\,F{d^3x\,d^3p\over d\nu\,d\Omega\,dA\,dt} =
	a^3 {h^4 \nu^3\over c^2} F.
	\label{JnufromF}
\end{equation}
Substituting $F$ from (\ref{JnufromF}) into (\ref{Feq}), we easily obtain:
\begin{equation}
	{\partial J_\nu\over\partial t} - 3{\dot{a}\over a}\,J_\nu +
	{\partial\over\partial x^i}\left(\dot{x}^i J_\nu\right) +
	{ a^3 h^4 \nu^3\over c^2} 
	{1\over p^2}{\partial\over\partial p}\left(p^2\dot{p} 
	{ c^2\over a^3 h^4 \nu^3} J_\nu \right) =
	{\rm RHS}(J_\nu),
	\label{Jnueqa}
\end{equation}
where we have written the divergence over momentum in spherical coordinates
explicitly presenting the derivative with respect to the total momentum
and noting that the derivatives with respect to orientation vanish
due to the isotropicity of the Robertson-Walker space-time.
Since $p=ah\nu/c$, we finally
obtain the following most general radiative transfer equation:
\begin{equation}
	{\partial J_\nu\over\partial t} +
	{\partial\over\partial x^i}\left(\dot{x}^i J_\nu\right) -
	H\left(\nu{\partial J_\nu\over\partial\nu}-3J_\nu\right) =
	- k_\nu J_\nu + S_\nu,
	\label{Jnueq}
\end{equation}
where $H$ is the Hubble constant, $H\equiv\dot{a}/a$, and we specified
physical processes including absorption with the coefficient $k_\nu$ and
emission (sources) $S_\nu$.

We now consider the mean specific intensity in the universe, averaged over
all space and all directions,
\begin{equation}
	\bar J_\nu(t) \equiv \langle J_\nu(t,x^i,n^k)\rangle,
	\label{Jnubardef}
\end{equation}
where the averaging operator $\langle\rangle$ acting on a function
$f(x^i,n^k)$ of position and direction is defined as:
\begin{equation}
	\langle f(x^i,n^k)\rangle = {\rm a~number} =
	\lim_{V\rightarrow\infty} {1\over 4\pi V}
	\int_V d^3x \int d\Omega f(x^i,n^k).
	\label{averagoper}
\end{equation}
Applying the averaging operator to the equation (\ref{Jnueq}), we obtain
the following equation for the mean specific intensity:
\begin{equation}
	{\partial \bar J_\nu\over\partial t} -
	H\left(\nu{\partial \bar J_\nu\over\partial\nu}-3\bar J_\nu\right) =
	- \bar k_\nu \bar J_\nu + \bar S_\nu,
	\label{Jnubareq}
\end{equation}
where, {\it by definition\/}, 
$$
	\bar S_\nu \equiv \langle S_\nu \rangle,
$$
and
$$
	\bar k_\nu \equiv {\langle k_\nu J_\nu\rangle\over \bar J_\nu}.
$$
We would like to stress here that we use the bar symbol 
``$\bar{\phantom{\ \ }}$'' merely as a part
of notation and not to denote the space average; in particular, 
$\bar k_\nu$ is not a space average of $k_\nu$ since it is averaged,
weighted
by the local value of specific intensity $J_\nu$.

The purpose of this section is, however, not to derive equation 
(\ref{Jnubareq}), but an equation for fluctuations in the specific intensity
around the mean. It is convenient therefore to introduce a new quantity
called relative (to the mean) specific intensity,
$f_\nu(t,x^i,n^k)$, as a measure of deviation of the specific intensity in
a particular position in space $x^i$ and in a particular direction $n^k$:
\begin{equation}
	J_\nu \equiv f_\nu\bar J_\nu,
	\label{fnudef}
\end{equation} 
so that $\langle f_\nu\rangle=1$. It is straightforward
to derive the following equation for $f_\nu$:
\begin{equation} 
	{\partial f_\nu\over\partial t} + 
	{\partial\over\partial x^i}\left(\dot{x}^i f_\nu\right) =
	H\nu{\partial f_\nu\over\partial\nu} - 
	(k_\nu-\bar k_\nu) f_\nu + \psi_\nu,
	\label{fnueqfull}
\end{equation}
where
\begin{equation}
	\psi_\nu \equiv {S_\nu - \bar S_\nu\over \bar J_\nu}.
	\label{psinudef}
\end{equation}
Equation (\ref{fnueqfull}) is tedious and by no means simpler than the
original equation (\ref{Jnueq}). However, if we restrict ourselves to
scales significantly smaller than the horizon size, and matter velocities
much smaller than the speed of light (Newtonian limit), then we can
simplify equation (\ref{fnueqfull}) substantially by noting that the
relative specific intensity does not change substantially and the universe
does not expand substantially over the period of time a photon needs to travel
over the scale under consideration. Therefore, we can ignore the time
derivative and the derivative with respect to the frequency.
Under these assumptions, equation (\ref{fnueqfull}) reduces to the following
simple equation:
\begin{equation}
	{c\over a} n^i{\partial f_\nu\over\partial x^i} = 
	- (k_\nu-\bar k_\nu) f_\nu + \psi_\nu
	\label{fnueq}
\end{equation}
Equations (\ref{Jnubareq}) and (\ref{fnueq}) are represent the radiative 
transfer equations in the expanding universe in the Newtonian limit.

\section{Local Optical Depth Approximation to the Radiative Transfer
Equations}

Our goal in this section is to describe numerical techniques for solving
equations (\ref{Jnubareq}) and (\ref{fnueq}) within the framework of
simulating large scale structure and galaxy formation in the
universe.
The former of those two
equations can be reduced to an ordinary differential equation and 
offers no difficulties in solving it (cf Gnedin 1996). The latter
one however is beyond practical implementation for the current
computer capabilities and resolution requirements for galaxy formation
simulations. 

The main difficulty comes from the fact that $f_\nu$ is
a function of six variables: three spatial coordinates and three
coordinates in the momentum space. In addition, unless $J_\nu$ by itself
is needed in a simulation, only direction-averaged quantities are
required to calculate photoionization and photoheating rates which are
needed to accurately compute gas temperature, ionization
fractions, cooling rates etc. But even if we approximate the directional
dependence of $f_\nu$, it still will be of little use since $f_\nu$
will remain a function of four arguments; storing such a function with
sufficient number of resolution elements in each of four direction
is a challenging problem for modern galaxy formation simulations.

The second complication with practical implementation of the radiative
transfer equation (\ref{fnueq}) is that the ionization state of cosmic
gas (and, therefore, the absorption coefficient $k_\nu$) 
can change on a very short time-scale, substantially shorter than the
hydrodynamic Courant time-scale. Equation (\ref{fnueq}) must therefore
be solved millions of times for a typical large cosmological simulation,
which is obviously well beyond existing computer power.

Fortunately, there exist two physical reasons why a reasonable
approximation to equation (\ref{fnueq}) can be found, that will employ
only functions of three arguments. 

There are two main physical effects which produce spatially variable
specific intensity: the optical depth of the gas, that can shield
high density regions from external radiations ($k_\nu$ is a function of
position), and nonuniformly distributed sources of radiation ($\psi_\nu$ is
a function of position). The optical depth is indeed a function of
seven arguments: the frequency $\nu$ and two spatial positions
(optical depth from one point to another); however, its dependence on
frequency is uniquely determined since for known column densities of
neutral hydrogen $N_{\HI}$, neutral helium $N_{\GI}$, and 
singly ionized helium 
$N_{\GII}$ (we consider
only hydrogen and helium contributions to the optical depth since heavy
element abundances are small) the optical depth is
\begin{equation}
	\tau_\nu \equiv N_{\HI} \sigma_\nu^{\HI} + N_{\GI}\sigma_\nu^{\GI} +
	N_{\GII}\sigma_\nu^{\GII},
	\label{taudefsig}
\end{equation}
where $\sigma_\nu^i$ is the photoionization cross-section for a species 
$i={\HI},{\GI},{\GII}$ and is a known function of frequency. In addition,
there is no need to know $\tau_\nu$ with high accuracy, since if the optical
depth is small, it does not matter if it is $10^{-3}$ or $10^{-5}$, and
if it is large, it again does not matter if it is $10^3$ or $10^5$. It is
only $\tau_\nu\sim1$ that should be computed accurately. However, since 
$\tau_\nu\sim1$ primarily happens at the boundary of an optically thick
region, the total volume occupied by regions with $\tau_\nu\sim1$ is
not large and little error in physical state of gas is introduced
from a large error in the optical depth calculation
\footnote{Those handwaving arguments can not serve as a proof to the
approximation described below but merely as justification for
searching for reasonable approximations.}.

The second important physical reason why equation (\ref{fnueq}) can 
be reasonably approximated with functions of positions only, is that
source function $\psi_\nu$ dependence on frequency, if known, does
not change with time since photons does not have enough time to redshift
significantly while traveling a distance short compared to the horizon
scale. Since $\psi_\nu$ includes only fluctuations in the source function,
it is mainly contributed to by a few nearby sources, while the redshift
of radiation emitted by distant (and, therefore, uniform) sources is
already taken into account by a derivative with respect to $\nu$ in
equation (\ref{Jnubareq}). Thus, the frequency dependence of relative
specific intensity $f_\nu$ at every particular point and at every moment 
in time can be easily calculated from scratch and needs not to be saved
as an additional dimension for $f_\nu$ (in particular, it is the same for
all positions).

Let us now proceed further to deriving the simplest possible approximation
that has any sense, which we call the Local Optical Depth Approximation.
First, let us consider the homogeneous equation (\ref{fnueq}) when no
spatial dependence of sources is taken into account. Let $f^{(0)}_\nu$
be its solution. It is convenient and
instructive to introduce optical depth $\tau_\nu$ as:
\begin{equation}
	f^{(0)}_\nu(t,x^i,n^k)\equiv \exp\left(-\left[\tau_\nu(t,x^i,n^k)-
	\bar \tau_\nu(t)\right]	\right)
	\label{taudef}
\end{equation}
where $\bar\tau_\nu(t)$ is defined as:
$$
	e^{-\bar\tau_\nu} \equiv \langle e^{-\tau_\nu}\rangle
$$
because $\langle f_\nu\rangle=1$. Substituting (\ref{taudef}) into
(\ref{fnueq}) with $\psi_\nu=0$ we obtain the following equation:
\begin{equation}
	n^i{\partial \tau_\nu\over\partial x^i} = 
	 {a\over c} (k_\nu-\bar k_\nu)
	\label{taunueq}
\end{equation}
with the formal solution (omitting the time dependency for clearness):
\begin{equation}
	\tau_\nu(x^i,n^k) = \tau^{(0)}_\nu(x^i-n^i(x^kn^k),n^k) +
	{a\over c} \int^0_{-(x^kn^k)} \left[ k_\nu(x^i+n^il)-\bar k_\nu
	\right] dl,
	\label{taunusol}
\end{equation}
where $\tau^{(0)}_\nu$ is an arbitrary function of two two-dimensional 
vectors which is specified from boundary conditions. 
One can see that $\tau_\nu$ is indeed an optical depth in a sense
that it is an integral of an absorption coefficient over a distance.

However, as we discussed before, there is no practical sense to implement
this equation in a real code, or even in a direction-averaged version of
the exact solution (in fact, it is not even clear how to calculate
the direction-averaged $f_\nu$ in a closed form from [\ref{taunusol}]).
More than that, what we would eventually like to get as an approximation to
$f_\nu$ is the completely local dependence of $f_\nu$ on properties of
the gas; this would allow us to consider cooling evolution within a 
hydrodynamic time-step of each resolution element independently and use
an ordinary differential equation solver to resolve smallest time-scales
in the problem (see Gnedin 1996 for details). We therefore introduce an
approximation to $f_\nu$ by defining $\tau_\nu$ in (\ref{taudef}) as
a function of position and frequency only by the following {\it ansatz\/}:
\begin{equation}
	\tau_\nu(x^i) \equiv \left[n_{\HI} \sigma_\nu^{\HI} + 
	n_{\GI}\sigma_\nu^{\GI} +
	n_{\GII}\sigma_\nu^{\GII}\right]L,
	\label{taulod}
\end{equation}
where $n_{\HI}$, $n_{\GI}$, and $n_{\GII}$ are local number densities of
neutral hydrogen, neutral helium, and singly ionized helium and 
$L$ is a characteristic distance (which is a function of position, but
not direction; the resulting $f^{(0)}_\nu$ is therefore a function of
position and frequency, but not direction). The justification to this
ansatz is that it is likely that ionization fractions are similar in the
neighboring points and if we choose the characteristic distance $L$
appropriately, we will obtain a successful approximation to $f_\nu$.
Since $L$ is obviously related to the characteristic scale over which
the density changes, we choose the following form for $L$:
\begin{equation}
	L = {\rho\over\sqrt{\alpha \left|\nabla\rho\right|^2 +
	\beta\rho\left|\Delta\rho\right|}},
	\label{defL}
\end{equation}
where parameters $\alpha=2.11$ and $\beta=0.27$ are fitted to produce 
the correct
value for the column density for a spherical density distribution with
inverse square law profile in two limits $r\rightarrow0$ and 
$r\rightarrow\infty$ (two constraints produce two parameter values);
these values for $\alpha$ and $\beta$ also give a very close match to
the gaussian density profile and for $r^{-4}$ (photoionized inverse
square law) profiles.

The Local Optical Depth approximation is uncontrolled approximation in a
sense that it seems impossible to define its range of applicability; we
can guarantee that it is not zero since it gives correct results for
a spherical density distribution with the inverse square law profile and
quite accurate results for other density profiles. In addition, we have 
tested the approximation by solving equation (\ref{fnueq}) numerically
for density distributions extracted at a few fixed epochs from simulations
discussed in this paper; the correspondence between the exact result and
the approximate $f_\nu$ is remarkably good. The epochs, however, have been 
chosen after the universe reionizes, so, $f_\nu$ deviates from unity 
appreciably only within isolated dense optically 
thick clouds and the Local Optical Depth 
approximation seems to give reasonable results in this limit; when the
most of the universe is ``optically thick'' (i.e.\ when $\bar\tau_\nu>1$)
the approximation is less accurate, but since $\tau_\nu>1$ in most of the
volume of the universe in that case, big errors in $\tau_\nu$ do not lead to
substantial errors in ionization abundances as have been argued above.

Now, equipped with the approximation to the solution $f^{(0)}$ for the 
homogeneous radiative transfer equation, we can obtain solution for
the whole equation (\ref{fnueq}). Since the direction $n^k$ enters
equation (\ref{fnueq}) only as a parameter, we can introduce a distance 
$l$ along the direction $n^k$ and equation (\ref{fnueq}) becomes an
ordinary differential equation, which can be easily solved
(we assume further that $\psi_\nu$ does not depend on direction,
i.e.\ that radiation sources are isotropic; in the opposite case 
dependence on direction can not be eliminated and the described
approximation becomes invalid):
\begin{equation}
	f_\nu(x^i,n^k) = e^{\displaystyle -\left[\tau_\nu(x^i)-\bar\tau_\nu
	\right]} +
	{a\over c}\int_0^\infty dl\, \psi_\nu(x^i+n^il)
	e^{\displaystyle -\tau_\nu(x^i,x^i+n^kl)},
	\label{fnusola}
\end{equation}
where we specify $\tau_\nu(x^i,x^i_1)$ below. Since no directional
dependence required in order to calculate effects of radiation on
the gas, we average equation (\ref{fnusola}) over all directions;
in order to keep notation compact, we retain the same symbol $f_\nu(x^i)$ for 
the direction averaged $f_\nu(x^i,n^k)$, i.e.
$$
	f_\nu(x^i) \equiv {1\over 4\pi} \int f_\nu(x^i,n^k) d\Omega.
$$
Then equation (\ref{fnusola}) becomes:
\begin{equation}
	f_\nu(x^i) = e^{\displaystyle -(\tau_\nu(x^i)-\bar\tau_\nu)} +
	{a\over c}\int d\Omega \int_0^\infty dl \psi_\nu(x^i+n^il)
	e^{\displaystyle -\tau_\nu(x^i,x^i+n^kl)}.
	\label{fnusolb}
\end{equation}
If we introduce a new vector $x_1^i\equiv x^i+n^kl$ and notice that
$l=|x^i-x_1^i|$ and $d^3x_1=l^2dl\,d\Omega$, we obtain the following
equation for $f_\nu(x^i)$:
\begin{equation}
	f_\nu(x^i) = e^{\displaystyle -(\tau_\nu(x^i)-\bar\tau_\nu)} +
	{a\over c}\int d^3x_1 { \psi_\nu(x_1^i)\over (x^i-x_1^i)^2}
	e^{\displaystyle -\tau_\nu(x^i,x_1^i)},
	\label{fnusolc}
\end{equation}
where $\tau_\nu(x^i,x_1^i)$ is the true optical depth from point $x^i$ to 
point $x_1^i$. 

Integral (\ref{fnusolc}), except for the exponential factor, is a potential
for the force with $1/r^3$ force law, which can be solved for by
known numerical techniques, for example, P$^3$M. The exponential factor,
however, offers a considerably difficulty in solving integral (\ref{fnusolc})
faster than in $O(N^2)$ operations.
We, therefore, leave this form of the solution to the radiative transfer 
equation (\ref{fnueq}) until further work.

\section{Molecular Hydrogen}

In this section we describe our method for solving for nonequilibrium
time-dependent molecular hydrogen abundance. The method we adopt is
exact but assumes that molecular hydrogen abundance is small compared
to the neutral hydrogen abundance and, thus, we can follow the ionization
balance without taking molecular hydrogen formation into account.

\placefig{\tabletwo}

We use Shapiro \& Kang\markcite{SK87} (1987) paper 
as our source of reaction rates.
The Table \ref{tabmolhyd} contains all relevant reactions together with
their number from Shapiro \& Kang\markcite{SK87} (1987) paper; we use this number as
an identification for a reaction.

Assuming that we know time-dependence of hydrogen and helium ion
abundances as a function of time, we can write down the following three
equations for three chemical species under consideration: $\MH$, $\Hp$,
and $\Hm$:
\begin{mathletters}
\begin{eqnarray}
	{1\over n_b}{dX_{\Hm}\over dt} & = & -X_{\Hm} 
	\left(k_8X_{\HI}+k_{16}X_e+k_{17}X_{\HI}+k_{18}X_{\HII}+
	k_{19}X_{\HII}+k_{21}X_{\Hp}+p_{27}\right) + \nonumber \\
	& & \left(k_7X_{\HI}X_e+k_{13}X_{\MH}X_e\right),
\end{eqnarray}
\begin{eqnarray}
	{1\over n_b}{dX_{\Hp}\over dt} & = & -X_{\Hp} 
	\left(k_{10}X_{\HI}+k_{20}X_e+k_{21}X_{\Hm}+
  	p_{28}+p_{30}\right) + \nonumber \\
	& & \left(k_9X_{\HI}X_{\HII}+k_{12}X_{\MH}X_{\HII}+
	k_{19}X_{\HII}X_{\Hm}+p_{29}X_{\MH}\right),
\end{eqnarray}
and
\begin{eqnarray}
	{1\over n_b}{dX_{\MH}\over dt} & = & -X_{\MH} 
	\left(k_{11}X_{\HI}+k_{12}X_{\HII}+k_{13}X_e+k_{14}X_e+
  	k_{15}X_{\MH}+p_{29}+p_{31}\right) + \nonumber \\
	& & \left(k_8X_{\HI}X_{\Hm}+k_{10}X_{\HI}X_{\Hp}+
	k_{21}X_{\Hp}X_{\Hm}\right),
\end{eqnarray}
where
\label{molhydeqsa}
\end{mathletters}
the fractional abundance $X_j$ for a species $j$ is defined as:
\begin{equation}
	n_j\equiv X_j {\rho\over m_H},
	\label{Xdef}
\end{equation}
where $m_H$ is the mass of a hydrogen atom and $\rho$ is the gas mass
density; the total baryon number density $n_b$ is
\begin{equation}
	n_b\equiv {\rho\over m_H},
	\label{nbdef}
\end{equation}
and temperature and $X_j$ for $j=\HI,\HII,\GI,\GII,\GIII,e$ are
known functions of time.

Equations \ref{molhydeqsa} have disadvantage that they are nonlinear,
and, therefore, do not offer a simple solution. The nonlinearity
comes in with two terms: $k_{15}X_{\MH}^2$ and $k_{21}X_{\Hp}X_{\Hm}$. 
However, there is a 
lucky coincidence that each of those terms is superseded by another term 
in each of equations (\ref{molhydeqsa}):
\begin{enumerate}
 \item $k_{21}X_{\Hp}X_{\Hm}\ll k_{18}X_{\HII}X_{\Hm}$ in equation for
	$\Hm$ since $k_{21}\approx k_{18}$;
 \item $k_{21}X_{\Hp}X_{\Hm}\ll k_{20}X_{\Hp}X_e$ in equation for
	$\Hp$ since $k_{21}\sim k_{20}$;
 \item $k_{21}X_{\Hp}X_{\Hm}\ll k_8X_{\HI}X_{\Hm}$ and 
	$k_{15}X_{\MH}^2\ll k_{11}X_{\MH}X_{\HI}$ in equation for
	$\MH$ since $k_{21}\la 10^3k_8$ and $k_{15}\sim k_{11}$,
	provided that $X_{\Hp}\ll 10^{-3}X_{\HI}$ (this condition
	is always satisfied in simulations described in this paper).
\end{enumerate}
Neglecting two nonlinear terms, we get a system of three
{\it linear\/} ordinary differential equations which offers an
{\it analytical\/} solution.

Let us introduce a three-dimensional vector $Y$ as follows:
\begin{equation}
	Y = \left( 
	\begin{array}{c}
	X_{\Hm} \\ X_{\Hp} \\ X_{\MH}\\ 
	\end{array}
	\right).
	\label{ydef}
\end{equation}
Then equations (\ref{molhydeqsa}) can be written in a matrix form as
follows:
\begin{equation}
	\dot{Y} = -A\,Y + S,
	\label{molhydeqsb}
\end{equation}
where $A$ is a matrix,
\begin{equation}
	A = n_b\left(
	\begin{array}{ccc}
	D_1 + k_8X_{\HI} + k_{19}X_{\HII}
  	& 0 & -k_{13}X_e \\
	-k_{19}X_{\HII} &
	D_2 + k_{10}X_{\HI}&
	-k_{12}X_{\HII}-p_{29} \\
	-k_8X_{\HI} & -k_{10}X_{\HI} & 
	D_3+k_{12}X_{\HII}+k_{13}X_e+p_{29}\\
	\end{array}
	\right),
	\label{adef}
\end{equation}
$D$ is a vector,
\begin{equation}
	D = \left(
	\begin{array}{c}
	k_{17}X_{\HI}+k_{16}X_e+k_{18}X_{\HII}+p_{27}\\
	k_{20}X_e+p_{28}+p_{30} \\
	k_{11}X_{\HI}+k_{14}X_e+p_{31} \\
	\end{array}
	\right),
	\label{ddef}
\end{equation}
and $S$ is a vector,
\begin{equation}
	S = n_b \left(
	\begin{array}{c}
	k_7X_{\HI}X_e\\
	k_9X_{\HI}X_{\HII}\\
	0\\
	\end{array} \right).
	\label{sdef}
\end{equation}

Equation (\ref{molhydeqsb}) can be solved {\it analytically\/} 
in a time interval $\Delta t$ from $t_0$ to $t_1=t_0+\Delta t$ 
in which both $A$ and $S$ do not change significantly with
the $O(\Delta t^2)$ to obtain:
\begin{equation}
	Y(t_1) = e^{\displaystyle -\bar{A}\Delta t}Y(t_0) +
	\left(1-e^{\displaystyle -\bar{A}\Delta t}\right)\bar{A}^{-1}\bar{S},
	\label{molhydsol}
\end{equation}
where 
\begin{equation}
	\bar{A} \equiv {1\over \Delta t}\int_{t_0}^{t_1} A(t) dt,
	\label{abardef}
\end{equation}
and 
\begin{equation}
	\bar{S} \equiv {1\over \Delta t}\int_{t_0}^{t_1} S(t) dt.
	\label{sbardef}
\end{equation}
Equation (\ref{molhydsol}) then gives the molecular hydrogen and 
$\Hp$ and $\Hm$ fractions at the time moment $t_1$.

\placefig{\end{document}}

\clearpage

\tableone

\clearpage

\tabletwo

\clearpage

\newcounter{figurecap}
\setcounter{figurecap}{0}

\begin{center}
\bf Figure Captions
\end{center}

\refstepcounter{figurecap}
Fig.\ \thefigurecap---\label{figTF}\capTF

\refstepcounter{figurecap}
Fig.\ \thefigurecap---\label{figSF}\capSF

\refstepcounter{figurecap}
Fig.\ \thefigurecap---\label{figTI}\capTI

\refstepcounter{figurecap}
Fig.\ \thefigurecap---\label{figDA}\capDA

\refstepcounter{figurecap}
Fig.\ \thefigurecap---\label{figFZ}\capFZ

\refstepcounter{figurecap}
Fig.\ \thefigurecap---\label{figMF}\capMF

\refstepcounter{figurecap}
Fig.\ \thefigurecap---\label{figZD}\capZD

\refstepcounter{figurecap}
Fig.\ \thefigurecap---\label{figZE}\capZE

\refstepcounter{figurecap}
Fig.\ \thefigurecap---\label{figEM}\capEM

\refstepcounter{figurecap}
Fig.\ \thefigurecap---\label{figEF}\capEF

\refstepcounter{figurecap}
Fig.\ \thefigurecap---\label{figSS}\capSS

\refstepcounter{figurecap}
Fig.\ \thefigurecap---\label{figZM}\capZM

\refstepcounter{figurecap}
Fig.\ \thefigurecap---\label{figFM}\capFM

\refstepcounter{figurecap}
Fig.\ \thefigurecap---\label{figMV}\capMV

\end{document}